\documentclass[usenatbib,usegraphicx,useAMS]{mn2e}

\usepackage{natbib}

\usepackage{bez123}
\usepackage{calc}
\usepackage{curves}
\usepackage{ebezier}
\usepackage{epic}
\usepackage{eepic}
\usepackage{latexpix}
\usepackage{multiply}
\usepackage{pgf}
\usepackage{rotating}
\usepackage{xcolor}
\usepackage{url}

\usepackage{multirow}

\usepackage{times}
\usepackage{amsmath}
\usepackage{amssymb}
\usepackage{bbold}
\usepackage{float}

\newcommand{\mbi}[1]{\mbox{\boldmath$#1$}}

\newcommand{\mat}[1]{\mbox{\rm\bf #1}}
\newcommand{\lsim}[1]{\mbox{${\,\hbox{\hbox{$ < $}\kern -0.8em \lower 1.0ex\hbox{$\sim$}}\,}$}}
\newcommand{\gsim}[1]{\mbox{${\,\hbox{\hbox{$ > $}\kern -0.8em \lower 1.0ex\hbox{$\sim$}}\,}$}}

\newcommand{\dd}{{\rm d}}

\newcommand{\cc}{{\rm c}}

\newcommand{\kk}{{\rm K}}

\def\beqn{\vspace{2mm}
\begin{eqnarray}} 
\def\eeqn{\vspace{2mm} 
\end{eqnarray}}

 \def\la{\langle}

\begin{document}
\title[Non-Gaussian field statistics]{Non-Gaussian gravitational clustering field statistics}

\author[Kitaura, Francisco-Shu]{Francisco-Shu Kitaura\thanks{E-mail: francisco.shukitaura@sns.it, kitaura@usm.lmu.de}\\
 SNS, Scuola Normale Superiore di Pisa, Piazza dei Cavalieri, 7 -- 56126 Pisa -- Italy \\
 LMU, Department of Physics, Ludwig-Maximilians Universit\"at M\"unchen, Scheinerstr. 1 -- D-81679 Munich -- Germany 
}

\maketitle

\begin{abstract}
In this work we investigate the multivariate statistical description of the matter distribution in the nonlinear regime.
 We introduce the multivariate Edgeworth expansion of the lognormal distribution to model the cosmological matter field. Such a technique could be useful to generate and reconstruct three-dimensional nonlinear cosmological density fields  with the information of higher order correlation functions. We explicitly calculate the expansion up to third order in perturbation theory making use of the multivariate Hermite polynomials up to sixth order. The probability distribution function for the matter field includes at this level the two-point, the three-point and the four-point correlation functions.  We use the hierarchical model to formulate the higher order correlation functions based on combinations of the two-point correlation function.  This permits us to find compact expressions for the skewness and kurtosis terms of the expanded lognormal field which can be efficiently computed.  The method is, however, flexible to incorporate arbitrary higher order correlation functions which have analytical expressions.
 The applications of such a technique can be especially useful to perform weak-lensing or neutral hydrogen 21 cm line tomography, as well as to directly use the galaxy distribution or the Lyman-alpha forest to study structure formation.

\end{abstract}


\section{introduction}

The cosmological  matter distribution encodes the information of the origin of our Universe and the processes which lead to structure formation. A precise understanding and modeling of its statistics is crucial to extract the cosmological information from observational data  and to ultimately  understand cosmic evolution. 

The Universe is being scrutinized with unprecedented accuracy. 
Many excellent astronomical surveys have been launched  in the recent past and ongoing and upcoming projects are on the way to perform the most ambitious map of the Universe up-to-date.

Some of the most successful low-redshift galaxy catalogs are 
the 2dF Galaxy Redshift Survey\footnote{http://www.mso.anu.edu.au/2dFGRS/} \citep[][]{2003astro.ph..6581C}
and the Sloan Digital Sky Survey (SDSS)\footnote{http://www.sdss.org/} \citep[][]{2009ApJS..182..543A}.
Deep surveys like the Baryon Oscillation Spectroscopic Survey (BOSS)\footnote{http://www.sdss3.org/cosmology.php} \citep[][]{2009astro2010S.314S},
 the DEEP2 Survey \footnote{http://deep.berkeley.edu} \citep[][]{2003SPIE.4834..161D} and
the VIMOS VLT Deep Survey (VVDS) \footnote{http://www.oamp.fr/virmos/vvds.htm}
 \citep[][]{2004A&A...428.1043L} are being run.

On the other hand  the Canada-France-Hawaii Telescope Legacy Survey (CFHTLS)\footnote{http://www.cfht.hawaii.edu/Science/CFHLS/} \citep[][]{2006ApJ...647..116H}, the VISTA (Visible and Infrared Survey Telescope for Astronomy)\footnote{http://www.vista.ac.uk/}
and the PANoramic Survey Telescope And Rapid Response System (Pan-STARRS)\footnote{http://pan-starrs.ifa.hawaii.edu/public/} \citep[][]{2002AAS...20112207K}, and the planned Dark Energy Survey (DES)\footnote{https://www.darkenergysurvey.org/} \citep[][]{2005astro.ph.10346T} or the  Joint Dark Energy Mission (JDEM)\footnote{http://jdem.gsfc.nasa.gov/}  from NASA-DOE and  Dark UNiverse Explorer (DUNE) from CNES \citep[][]{2005astro.ph..7043C,2006SPIE.6265E..58R} will provide the first weak lensing surveys covering very large sky areas and depth. 

Also the Lyman alpha forest will be a useful observable to study cosmology using for instance the BOSS survey \citep[the matter power-spectrum has already been measured with the SDSS, see][]{2005ApJ...635..761M}.

The neutral hydrogen 21-cm line will provide a new astronomical window to study structure formation in the Universe. Some of the most notable projects are
the Giant Metre-wave Radio Telescope (GMRT)\footnote{http://gmrt.ncra.tifr.res.in/} \citep[][]{2008AIPC.1035...75P}, the Precision Array to Probe Epoch of Reionization (PAPER)\footnote{http://astro.berkeley.edu/~dbacker/eor/} \citep[][]{2010AJ....139.1468P}, the LOw Frequency ARray (LOFAR)\footnote{http://www.lofar.org/}  \citep[][]{2007HiA....14..386F} and the Murchison Widefield Array (MWA)\footnote{http://www.MWAtelescope.org/} \citep[][]{2009IEEEP..97.1497L}.

In summary, an avalanche of astronomical data is being collected to study its structure and history based on different observables. In order to extract valuable cosmological information not only a careful modeling of the systematics of the observation process and the nature of the observable is required, but also a precise modeling of the underlying signal. We focus here on the cosmological matter density field.

Different approaches can be found in the literature to reconstruct the large-scale structure. 
 Geometrical reconstruction methods try to approximately capture the higher order statistics beyond the two-point correlation function in an effective way through a geometry-based prescription to form structures from a point source distribution.
The salient and pervasive foamlike pattern of the cosmic web has led to develop  methods like the Voronoi or Delaunay tessellations  \citep[see for example][]{2009arXiv0912.3448V,2007A&A...474..315A,2000A&A...363L..29S}.
On the other hand one can find reconstruction methods based on a statistical approach. 
The advantage of the statistical methods with respect to the geometrical ones is that one can clearly specify the assumptions made on the matter field and the observable in form of probability distribution functions. The statistical  methods could be more suitable to extract statistical quantities like the power-spectrum. This has been succesfully done in the Cosmic Microwave Background (CMB) for which the fluctuations can be assumed to be Gaussian distributed \citep[][]{2007ApJ...656..641E}.
 The disadvantage is that the level of complexity that such methods can achieve has always been limited as the formulation of the probability distribution functions and its applications to reconstruction methods is relatively complex and computer-intensive. Indeed, any complex realistic formulation seemed to be untreatable.
For a long time the Wiener-filter, also-called least-squares filter, has been the only available method in Astronomy to incorporate the statistical information in the reconstruction method \citep[see for example][]{1994ApJ...432L..75B,zaroubi,1995MNRAS.272..885F,1997MNRAS.287..425W,1999ApJ...520..413Z,2004MNRAS.352..939E,2006MNRAS.373...45E}.
Less attention was paid in the astrophysics community to the nonlinear version of the least-squares filter proposed by \citet[][]{tarantola}. Here the data model which relates the measurements to the seeked signal is extended to be nonlinear. Probably the first group applying this method in an astrophysical context was \citet[][]{1998A&A...336..137C} to  map the interstellar absorption structures in the galactic plane. \citet[][]{pichon} proposed to use this nonlinear reconstruction scheme to recover the cosmic density and velocity field traced by the Lyman alpha forest. 
The drawback of \citet[][]{tarantola}'s  approach is that it requires both a Gaussian prior  (with a nonlinear transformation) and a Gaussian likelihood for the distribution of the observable. This is a too crude assumption for many observables, like for instance a galaxy distribution. Recently the Poisson-lognormal (and Gaussian-lognormal) model was proposed in a Bayesian framework to recover the cosmic density field  \citep[][]{kitaura_log}. In this study it was shown that the lognormal prior is in good agreement with the underlying matter field extracted from  N-body simulations in the large overdense regions ($>10^3$), but fails to fit the matter statistics in the underdense regions. One of the  advantages of this method with respect to the nonlinear least-squares approach is that it can deal with non-Gaussian likelihoods.
Another important point of the Bayesian approach proposed in \citep[][]{kitaura_log} is that it can be easily extended to sample full posterior distributions \citep[see the works by][]{jasche_hamil,jasche_sdss,kitaura_lyman}. 

Nevertheless, non of the above mentioned statistical methods  includes any information beyond the two-point correlation function. As gravitational clustering forms nonlinear structures the Universe becomes inhomogenous and complex patterns arise which encode high order statistics.

The purpose of this work is to extend the multivariate characterization of matter  beyond the two-point correlation function to incorporate higher order statistics.
 We therefore relax the lognormal assumption and introduce the multivariate Edgeworth expansion which leads to additional terms describing the skewness and kurtosis of the field with respect to the lognormal distribution function.  
  This work is based on the univariate Edgeworth expansion introduced by \citet[][]{1991ApJ...381..349S,1995ApJ...442...39J,1995ApJ...443..479B} and \citet[][]{colombi}.
The Edgeworth expansion we find deviates from the trivial generalization of the univariate case to the multivariate case.
We use the hierarchical model \citep[see][]{1978ApJ...221...19F,1984ApJ...277L...5F,1986ApJ...306..358F,1989A&A...220....1B} to formulate the three-point and four-point correlation functions which permits us to find particular expressions for the skewness and kurtosis terms. The expressions we find are compact due to the symmetries introduced by the hierarchical model and can be computed by means of convolutions with fast Fourier transforms (fft's).
As the hierarchical model may fail at certain scales and regimes \citep[][]{1994ApJ...420..504S,1994ApJ...420..497M} this work could be extended incorporating more complex higher order correlation functions which include galaxy biasing or redshift distortions \citep[see the works by][]{1998ApJ...496..586S,1999ApJ...522...46T,2003ApJ...584....1M,2004ApJ...614..527Z,2008PhRvD..78h3519M} and to perform topological and morphological  studies \citep[see for example][]{1996ApJ...463..409M,2008ApJ...675...16G,2009MNRAS.394..454J}.

We believe that the method introduced in this paper can be very  useful to study cosmological structures in the range between the quasi-nonlinear and the nonlinear regime. It could be interesting to apply higher order statistics to galaxy redshift surveys, to weak-lensing surveys, to the Lyman alpha forest or to the 21 cm line. We would like to warn the reader that this work is still in a development phase as higher order correlation models need to be tested and many numerical studies still have to be done. This is the first of a series of works in which we will analyze the statistical description of gravitational clustering.

This paper is structured as follows. In the next section the gravitational clustering statistics will be analyzed in great detail (section \ref{sec:GCstats}). We will start reviewing the work done so far for the univariate matter distribution (section \ref{sec:unicase}) and then present the multivariate case (section \ref{sec:MV}). This will lead us to a multivariate Edgeworth expansion of the lognormal field up to third order in perturbation theory including two-point, three-point and four-point correlation functions. Then we will present the hierarchical model  (section \ref{sec:HM}) and use the expression for the three-point  correlation function to calculate the skewness and kurtosis terms in the Edgeworth expansion (section \ref{sec:edgeHM}). A detailed calculation can be found in the appendix.  Finally we will present the summary and conclusions of this work. 

\pagebreak

\section{Gravitational clustering field statistics}

\label{sec:GCstats}

In this section we will study the matter field statistics produced by gravitational clustering. We start with a physical motivation followed by the review of the univariate non-Gaussian statistics. Then we introduce the multivariate Edgeworth expansion of the Lognormal field. Finally we present the hierarchical model and calculate  the skewness and kurtosis terms of the Edgeworth expansion.

\subsection{Physical motivation}
\label{sec:mot}

Let us divide the Universe into $N_{\rm c}$ cells and assign to each cell $i$ a position $r_i$, a matter density $\rho_i$ and a peculiar  velocity $v_i$.
 The continuity equation relates the evolution of the matter content in the Universe to its peculiar velocity field:
\begin{eqnarray}
\label{eq:cont}
\hspace{2cm}\frac{{\partial}\rho(\mbi r)}{\partial t}+\frac{1}{a}\nabla_{\mbi r}\cdot(\rho(\mbi r)\mbi v)&=&0\nonumber \\
\hspace{2cm}\frac{{\rm d}\mbi\rho}{{\rm d}t}+\frac{1}{a}\rho(\mbi r)\nabla_{\mbi r}\cdot\mbi v&=&0 {,}
\end{eqnarray}
with  $t$ being the cosmic time, $a$ the scale factor, $\mbi r$ the set of positions ($\{r_1,\dots,r_{N_{\rm c}}\}$), $\mbi\rho$ the matter density field ($\{\rho_1,\dots,\rho_{N_{\rm c}}\}$), $\mbi v$ the peculiar velocity field ($\{v_1,\dots,v_{N_{\rm c}}\}$) and ${\rm d}/{\rm d} t=\partial /\partial t+1/a\,(\mbi v \cdot\nabla_{\mbi r})\rho(\mbi r)$ the total derivative.

We can follow matter particles until they start crossing-over (in this regime particles can have different peculiar velocities at the same position) and form caustics. Before this occurs we can write the formal solution to Eq.~(\ref{eq:cont}) as:
\begin{equation}
  \label{eq:solcont}
  \mbi\rho=\langle\mbi\rho\rangle e^{\mbi s} ,\,\,\, {s}=-\int \dd t \frac{ 1 }{ a }\nabla_{\mbi r} \cdot \mbi v {,} 
\end{equation} 
with $\mbi s$ being the logarithm of the normalized density:
\begin{equation}
  \label{eq:logdens}
  \mbi s\equiv\ln \mbi \rho-\ln\langle\mbi \rho \rangle=\ln (\mbi \rho/\langle\mbi \rho \rangle)=\ln(1+\mbi \delta_{{\rm M}}){,} 
\end{equation} 
and the matter overdensity field given by: $\mbi \delta_{{\rm M}}=\mbi \rho/\langle\mbi\rho \rangle-\vec{1}$. The ensemble averages are used  at this stage only to denote the mean of the variable.
Note that the field $\mbi s$ does  not have zero mean, but is given by: 
$\mbi \mu_{s}\equiv\langle \mbi s \rangle=\langle\ln\mbi\rho \rangle-\ln\langle\mbi \rho \rangle$.
It is convenient to define a field $\mbi \Phi$ with zero mean ($\langle\mbi\Phi\rangle=0$):   
\begin{equation}
  \label{eq:phi}
  \mbi \Phi\equiv \ln \mbi \rho- \langle\ln \mbi\rho \rangle=\mbi s-\mbi\mu_{ s} \,.  
\end{equation} 
 Assuming that $\mbi \Phi$ is a Gaussian random field leads to a lognormal distributed  density field \citep[see][]{1991MNRAS.248....1C}. Note however, that Lagrangian perturbation theory -which is known to give a good approximation of gravitational clustering until shell crossing starts \citep[see e.g.][]{1993MNRAS.264..375B,1994MNRAS.267..811B,bouchet1995}- deviates from the lognormal distribution already in the linear \citet[][]{1970A&A.....5...84Z} approximation  \citep[see][]{1993ApJ...410..482P,1995ApJ...443..479B}. Furthermore, after structures start to virialize the peculiar velocity field will be strongly modified and Lagrangian perturbation theory will start to fail dramatically. 
\citet[][]{colombi} suggested to use the formalism developed by \citet[][]{1995ApJ...442...39J} and \citet[][]{1995ApJ...443..479B} to study the departures from the lognormal distribution function including higher order correlation functions in the univariate matter distribution  \citep[see also the work on a generalized lognormal distribution by][]{2000MNRAS.313..725S}. 

\subsection{Univariate case}
\label{sec:unicase}

In this subsection we will revise the matter statistics for the one-dimensional probability distribution function as developed in the works by \citet[][]{1995ApJ...442...39J,1995ApJ...443..479B} and \citet[][]{colombi}. For a general overview on asymptotic statistical techniques see  \citet[][]{Berkowitz:1970:CMH} and  \citet[][]{barndorff}. 
Other univariate matter field distribution functions have been proposed. They are however not trivially extendable to the multivariate case. Either they do not include higher order correlations, but are extracted from fitting the univariate matter distribution based on numerical N-body simulations \citep[e.g.~][]{2000ApJ...530....1M}, or a distribution function function is expanded using  the variance as an univariate parameter \citep[]{2000ApJ...539..522G}. For this reason we will consider only the above mentioned approach based on the expansion of the lognormal distribution function.

Let us define the quantity $\nu$ with zero mean and unity variance:
\begin{equation}
\label{eq:nu}
  \nu\equiv \sigma^{-1} \Phi \,,  
\end{equation} 
with $\sigma^2=\langle \Phi^2\rangle$ being the variance of  $\Phi$ (the relation between  the variance of $\Phi$ and the variance of the matter overdensity $\delta_{\rm M}$ is derived in appendix \ref{app:univar}).
Higher order moments of $\nu$ can be found by calculating the ensemble average of powers of $\nu$ over the probability distribution function $P(\nu)$:
\begin{equation}
\mu_n \equiv \int \dd \nu P(\nu) \,\nu^n=\langle\nu^n\rangle { , }
\end{equation} 
with $n$ being the order of the moment.
Please note that the moment $\mu$ refers to the variable $\nu$ and not to the variable $s$.
The moment generating function is given by:
\begin{equation}
  \label{eq:momgenfunc}
  \mathcal M_\nu(t)\equiv \sum_{n=0}^\infty \mu_n\frac{t^n}{n!} =\int \dd\nu P(\nu) e^{t\nu}=\langle e^{t\nu}\rangle\,.
\end{equation} 
Subsequent derivatives of $\mathcal M_\nu(t)$ at the origin $t=0$ yield the moments:
\begin{equation}
  \mu_n =\frac{\dd^n\mathcal M_\nu(t)}{\dd t^n}\bigg|_{t=0}{ , }
\end{equation} 
The cumulant generating function is given by:
\begin{equation}
  \mathcal C(t)\equiv \sum_{n=1}^\infty \kappa_n \frac{t^n}{n!}{ , }
\end{equation} 
with $\kappa_n$ being the cumulants or connected moments: 
\begin{equation}
  \kappa_n \equiv\langle \nu^n\rangle_\cc {.}
\end{equation} 
The cumulants can be obtained from the relation between the moment and cumulant generating functions \citep[see for example][]{2002PhR...367....1B}:
\begin{equation}
  \label{eq:momcum}
  \mathcal M_\nu(t)=\exp(\mathcal C(t)) {,}
\end{equation} 
or equivalently: $\mathcal C(t)=\ln \left(\mathcal M_\nu(t)\right)$. 
Hence, the cumulants can be calculated by: 
\begin{equation}
  \kappa_n =\frac{\dd^n\ln\left(\mathcal M_\nu(t)\right)}{\dd t^n}\bigg|_{t=0}{.}
\end{equation} 
It is however, more convenient to use expression (\ref{eq:momcum}) to relate the moments to the cumulants:  
\begin{equation}
 \sum_{n=0}^\infty \frac{1}{n!}\mu_n t^n=\exp\left( \sum_{n=1}^\infty \frac{1}{n!}\kappa_n t^n \right) {.}
\end{equation} 
This equation yields for the first order moments:
\begin{eqnarray}
\label{eq:hemr1D}
\mu_0&=&1\\
\mu_1&=&\kappa_1=0\nonumber\\
\mu_2&=&\kappa_2+\kappa_1^2=1\nonumber\\
\mu_3&=&\kappa_3+3\kappa_2\kappa_1+\kappa_1^3=\kappa_3\nonumber\\
\mu_4&=&\kappa_4+4\kappa_3\kappa_1+3\kappa_2^2+6\kappa_2\kappa_1^2+\kappa_1^4=\kappa_4+3\nonumber\\
\mu_5&=&\kappa_5+5\kappa_4\kappa_1+10\kappa_3\kappa_2+10\kappa_3\kappa_1^2+15\kappa_2^2\kappa_1+10\kappa_2\kappa_1^3+\kappa_1^5\nonumber\\
&=&\kappa_5+10\kappa_3\nonumber\\
\mu_6&=&\kappa_6+6\kappa_5\kappa_1+15\kappa_4\kappa_2+15\kappa_4\kappa_1+10\kappa_3^2+60\kappa_3\kappa_2\kappa_1\nonumber\\
&&+20\kappa_3\kappa_1^3+15\kappa_2^3+45\kappa_2^2\kappa_1^2+15\kappa_2\kappa_1^4+\kappa_1^6\nonumber\\
&=&\kappa_6+15\kappa_4+10\kappa_3^2+15\nonumber\,.
\end{eqnarray}

The probability distribution function of $\nu$ can be obtained by inverting Eq.~(\ref{eq:momgenfunc})
using the inverse Laplace transform  \citep[see the review by][]{2002PhR...367....1B}: 
\begin{equation}
P(\nu)=\int^{\sqrt{-1}\infty}_{-\sqrt{-1}\infty}\frac{\dd t}{2\pi\sqrt{-1}}\exp\left(t\nu+\mathcal C(t)\right) {.}
\end{equation} 
Thus, the generating function fully defines the probability distribution function. 
When the departures from the Gaussian distribution function are small one can expand $P(\nu)$ in a  Gram-Charlier series:
\begin{equation}
P(\nu)=G(\nu)\left[1+ \sum^\infty_{l=1} \frac{1}{l!}c_l (-1)^l h_l(\nu)\right] {,}
\end{equation} 
with $G(\nu)=\frac{1}{\sqrt{2\pi}}e^{-\frac{\nu^2}{2}}$ and $h_l(\nu)$ being the Hermite polynomials.
The Hermite polynomial of degree $n$ can be calculated by:
\begin{equation}
\label{eq:hermform}
h_n(\nu)\equiv(-1)^n e^{\frac{1}{2}\nu^2}\frac{{\rm d}^n}{{\rm d}\nu^n}e^{-\frac{1}{2}\nu^2} {.}
\end{equation}

This leads to the following first polynomials:
\begin{eqnarray}
\label{eq:hemr1D}
h_0(\nu)&=&1\\
h_1(\nu)&=&\nu\nonumber\\
h_2(\nu)&=&\nu^2-1\nonumber\\
h_3(\nu)&=&\nu^3-3\nu\nonumber\\
h_4(\nu)&=&\nu^4-6\nu^2+3\nonumber\\
h_5(\nu)&=&\nu^5-10\nu^3+15\nu\nonumber\\
h_6(\nu)&=&\nu^6-15\nu^4+45\nu^2-15\nonumber\,.
\end{eqnarray}
Using the orthogonality relation:
\begin{equation}
\label{eq:hermprop}
\int \dd \nu \,G(\nu)h_l(\nu)h_m(\nu)=\begin{cases}
 0 ,& \mbox{if $l\neq m$;}  \\
 l! ,& \mbox{otherwise,}
 \end{cases}
\end{equation}
we can calculate the Gram-Charlier coefficients $c_l$:
\begin{equation}
c_l=(-1)^l \int \dd \nu P(\nu)\, h_l(\nu)=(-1)^l \langle h_l(\nu)\rangle {.}
\end{equation} 
The latter expression yields:
\begin{eqnarray}
\label{eq:gram1D}
c_0&=&1\\
c_1&=&0\nonumber\\
c_2&=&0\nonumber\\
c_3&=&-\mu_3=-\kappa_3\nonumber\\
c_4&=&\mu_4-3=\kappa_4\nonumber\\
c_5&=&-\mu_5+10\mu_3=-\kappa_5\nonumber\\
c_6&=&\mu_6-15\mu_4+30=\kappa_6+10\kappa_3^2\nonumber\,.
\end{eqnarray}
The Gram-Charlier series may have poor convergence properties \citep[see][]{cramer}, and in some cases even violently diverge \citep[see][]{1998A&AS..130..193B}. For this reason \citet[][]{1995ApJ...442...39J} suggested to model the departures from the Gaussian distribution with an Edgeworth expansion which is a true asymptotic expansion \citep[for a full explicit expansion for arbitrary order see][and references therein]{1998A&AS..130..193B}. 
 The Edgeworth expansion consists of rearranging the terms in the Gram-Charlier series based on collecting all the terms with the same clustering strength. To see how to do this let us recall how the different terms scale with $\sigma$.

In perturbation theory the matter field is expanded as a sum of terms with increasing order: $\Phi=\Phi_1+\Phi_2+\Phi_3+\dots$, with $\Phi_1=\mathcal O(\sigma^1)$ being the linear term, $\Phi_2=\mathcal O(\sigma^2)$ being the quadratic term, $\Phi_3=\mathcal O(\sigma^3)$ being the cubic term, etc..
The linear term $\Phi_1$ is assumed to be Gaussian distributed, or equivalently in our case, the matter density field is assumed to be lognormal distributed at first order. 
To calculate the subsequent perturbation terms one needs to use higher order correlation functions.
As it was shown by \citet[][]{1984ApJ...277L...5F} the cumulants of order higher than two scale as: $\kappa_n=\mathcal O(\sigma^{2n-2})$. For this reason the following normalized cumulant  has  usually been introduced to calculate the Edgeworth expansion: 
\begin{equation}		
\label{eq:sn}
S_n\equiv \frac{\langle \nu^n\rangle_\cc}{\sigma^{2n-2}}\,,
\end{equation}
with $S_3=\frac{\langle \nu^3\rangle}{\sigma^{4}}$ and $S_4=\frac{\langle \nu^4\rangle-3\sigma^{4}}{\sigma^{6}}$.
Using the latter expression one can group the terms with the same  scaling power of $\sigma$.  
The Edgeworth expansion until third order perturbation is then given by:
\begin{eqnarray}
\label{eq:edge1V}
\lefteqn{P(\Phi)=\frac{\dd\nu}{\dd\Phi}\cdot P(\nu)=\sigma^{-1}\frac{1}{\sqrt{2\pi}}\exp\left(-\frac{(\sigma^{-1}\Phi)^2}{2}\right)}\nonumber\\
&&\times\left[1+\sigma\left(\frac{1}{3!}  S_3  h_3(\sigma^{-1}\Phi)\right)\right.\nonumber\\
&&\left.+\sigma^2\left(\frac{1}{4!} S_4 h_4(\sigma^{-1}\Phi)+\frac{10}{6!} S_3^2 h_6(\sigma^{-1}\Phi)\right)+\dots\right]\,.
\end{eqnarray}

\begin{figure*}				
\begin{center}
\vspace{0cm}
\input{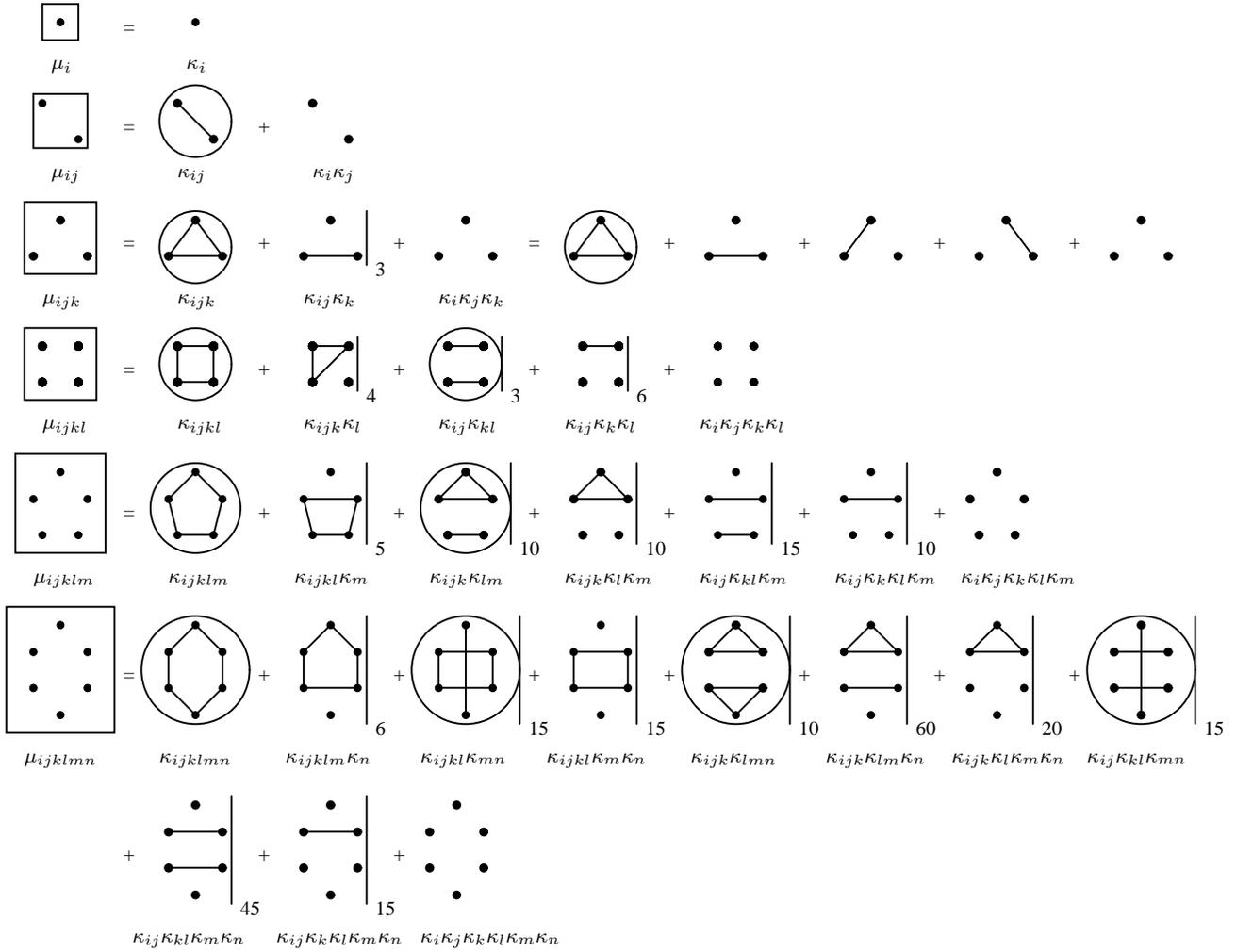}
\caption{Moments up to six point correlation statistics. The rectangular box stands for the ensemble average over the points, i.e. the moments. The points connected through lines represent the connected moments. In case of more than one equivalent element the number of elements is indicated.  The circles mark the terms which are not automatically zero as they do not contain a single unconnected point (we deal here with a centered variable with zero mean). The three equivalent terms for the third order moment have been specified. Let us define a cumulant object as a set of connected points in a term of a moment. Two cumulant objects will be equivalent if they have the same number of points. The number of equivalent elements $N_{\rm e}$ in a term is calculated by the factorial of the order of the moment $n$ divided by the factorial of the number of single points $N_{{\rm p}0}$, the factorial of the number of equivalent cumulant objects $N_{\rm obj}$ in a term and the factorial of the number of points for each cumulant object $\prod_k N_{{\rm p}k}$: $N_{\rm e}={n!}/({N_{{\rm p}0}!N_{\rm obj}!\prod_k N_{{\rm p}k}!})$. As an example the eighth term of the sixth order moment: $\kappa_{ij}\kappa_{kl}\kappa_{mn}$ has 15 equivalent elements: $15={6!}/({3!2!2!2!})$. }
\label{fig:cum}
\end{center}
\end{figure*}

\subsection{Multivariate case}
\label{sec:MV}

In this subsection we generalize the relations of the univariate matter distribution to the multivariate case. 
Now $\mbi \Phi$, $\mbi \rho$ and $\mbi s$ are scalar fields:   
\begin{equation}
  \Phi_i\equiv \ln \rho_i- \langle\ln \rho \rangle=s_i-\mu_{s} \,,  
\end{equation} 
and 
\begin{eqnarray}
  \lefteqn{S_{ij}\equiv\langle\Phi_i\Phi_j\rangle= \langle(s_i-\mu_s)(s_j-\mu_s)\rangle}\\
  &&=\langle\ln(1+\delta_{{\rm M}i})\ln(1+\delta_{{\rm M}j})\rangle-\langle\ln(1+\delta_{{\rm M}i})\rangle\langle\ln(1+\delta_{{\rm M}j})\rangle \nonumber\,,  
\end{eqnarray} 
with $\mat S$ being the variance of the field $\ln(1+\mbi\delta_{{\rm M}})-\mu_s$ (see appendix \ref{app:multivar} for the relation between  the variance of $\mbi\Phi$ and the variance of the matter overdensity field $\mbi\delta_{\rm M}$).
We introduce the field $\mbi \nu$ which has zero mean and unity variance by definition:
\begin{equation}
  \nu_i\equiv \sum_j S_{ij}^{-1/2} \Phi_j \,.
\end{equation} 

The $n$-dimensional moments are given by:
\begin{equation}
\mu_{i_1\dots i_n} \equiv \int \dd \mbi\nu P(\mbi\nu)\, \nu_{i_1}\dots\nu_{i_n}  =\langle\nu_{i_1}\dots\nu_{i_n}\rangle\,.
\end{equation} 
The multivariate moment generating function yields:
\begin{eqnarray}
\label{eq:momMV}
\mathcal M_{\mbi \nu}(t_{1}\dots t_{n})&\equiv& \sum_{q_{1}\dots q_{n}=0}^1 \langle \nu_{i_1}^{q_{1}}\dots \nu_{i_n}^{q_{n}} \rangle  \frac{t_{1}^{q_{1}}\dots t_{n}^{q_{n}}}{q_{1}!\dots q_{n}!}\nonumber\\
&=&\langle \exp\left(\sum_{l} t_{l}\nu_{i_l}\right)\rangle\,.
\end{eqnarray} 
Analogously to the univariate case subsequent derivatives of $\mathcal M_{\mbi \nu}(\mbi t)$ at the origin $\mbi t=0$ lead to the moments:
\begin{equation}
\mu_{i_1\dots i_n} =\frac{\partial^n \mathcal M_{\mbi \nu}(t_{1}\dots t_{n})}{\partial t_{1}\dots \partial t_{n}}\bigg|_{t_{1}\dots t_{n}=0}\,.
\end{equation} 
The cumulant generating function is given by:
\begin{equation}
\label{eq:cumMV}
\mathcal C(t_{1}\dots t_{n})\equiv \sum_{q_{1}\dots q_{n}=0}^1\langle \nu_{i_1}^{q_{1}}\dots \nu_{i_n}^{q_{n}} \rangle_\cc  \frac{t_{1}^{q_{1}}\dots t_{n}^{q_{n}}}{q_{1}!\dots q_{n}!}\,,
\end{equation} 
with $\kappa_{i_1\dots i_n}$ being the cumulants or connected moments:
\begin{equation}
\kappa_{i_1\dots i_n} =\langle \nu_{i_1}\dots \nu_{i_n} 	\rangle_\cc\,.
\end{equation} 
The moments are related to the cumulant generating functions by:
\begin{equation}
\mathcal M_{\mbi \nu}(t_{1}\dots t_{n})=\exp(\mathcal C(t_{1}\dots t_{n}))\,.
\end{equation}  
Plugging in Eqs.~(\ref{eq:momMV}) and (\ref{eq:cumMV}) in the latter expression we obtain:
\begin{eqnarray}
\hspace{-.2cm}\lefteqn{\sum_{q_{1}\dots q_{n}=0}^1 \langle \nu_{i_1}^{q_{1}}\dots \nu_{i_n}^{q_{n}} \rangle  \frac{t_{1}^{q_{1}}\dots t_{n}^{q_{n}}}{q_{1}!\dots q_{n}!} =}\nonumber\\
&&\exp\left( \sum_{q_{1}\dots q_{n}=0}^1\langle \nu_{i_1}^{q_{1}}\dots \nu_{i_n}^{q_{n}} \rangle_\cc \frac{t_{1}^{q_{1}}\dots t_{n}^{q_{n}}}{q_{1}!\dots q_{n}!} \right)\,.
\end{eqnarray} 
A way to spare the tedious calculations consists on looking at all combinations of connections between points in a diagram \citep[see Fig.~\ref{fig:cum} and][]{2002PhR...367....1B}.

Since the mean of $\nu_i$ is zero ($\mu_i=0$) we got rid off all the terms containing single connected points ($\kappa_i$). The result for the first moments is listed below: 
\begin{eqnarray}
\label{eq:hemrMV}
\mu_0&=&1\\
\mu_i&=&\kappa_i=0\nonumber\\
\mu_{ij}&=&\kappa_{ij}=\delta^\kk_{ij}\nonumber\\
\mu_{ijk}&=&\kappa_{ijk}\nonumber\\
\mu_{ijkl}&=&\kappa_{ijkl}\nonumber\\
&&\hspace{-1.cm}+\left[\frac{1}{2^3}\sum_{j_1\dots j_4\in[1,\dots,4]}\tilde{\epsilon}_{j_1\dots j_4}\delta^{\rm K}_{i_{j_1}i_{j_2}}\delta^{\rm K}_{i_{j_3}i_{j_4}}\right]_{3=\frac{4!}{2^3}}\nonumber\\
\mu_{ijklm}&=&\kappa_{ijklm}\nonumber\\
&&\hspace{-1.cm}+\left[\frac{1}{3!2}\sum_{j_1\dots j_5\in[1,\dots,5]}\tilde{\epsilon}_{j_1\dots j_5}\kappa_{i_{j_1}i_{j_2}i_{j_3}}\delta^{\rm K}_{i_{j_4}i_{j_5}}\right]_{10=\frac{5!}{3!2}}\nonumber\\
\mu_{ijklmn}&=&\kappa_{ijklmn}\nonumber\\
&&\hspace{-1.cm}+\left[\frac{1}{4!2}\sum_{j_1\dots j_6\in[1,\dots,6]}\tilde{\epsilon}_{j_1\dots j_6}\kappa_{i_{j_1}i_{j_2}i_{j_3}i_{j_4}}\delta^{\rm K}_{i_{j_5}i_{j_6}}\right]_{15=\frac{6!}{4!2}}\nonumber\\
&&\hspace{-1.cm}+\left[\frac{1}{3!3!2}\sum_{j_1\dots j_6\in[1,\dots,6]}\tilde{\epsilon}_{j_1\dots j_6}\kappa_{i_{j_1}i_{j_2}i_{j_3}}\kappa_{i_{j_4}i_{j_5}i_{j_6}}\right]_{10=\frac{6!}{3!3!2}} \nonumber\\
&&\hspace{-1.cm}+\left[\frac{1}{2^3}\sum_{j_1\dots j_6\in[1,\dots,6]}\tilde{\epsilon}_{j_1\dots j_6}\delta^{\rm K}_{i_{j_1}i_{j_2}}\delta^{\rm K}_{i_{j_3}i_{j_4}}\delta^{\rm K}_{i_{j_5}i_{j_6}}\right]_{15=\frac{6!}{2^3}}\nonumber \,,
\end{eqnarray}
where we have used the following identities: $i\equiv i_1,j\equiv i_2,k\equiv i_3,l\equiv i_4,m\equiv i_5,n\equiv i_6$
, the Kroenecker delta: $\delta^{\rm K}$ and the modified Levi-Civita tensor we introduce here: 
\begin{equation}
\tilde{\epsilon}_{i_{j_1}\dots i_{j_n}}\equiv(-1)^{N_{\rm t}}\epsilon_{i_{j_1}\dots i_{j_n}}\,,
\end{equation}
with $N_{\rm t}$ being the number of transpositions. This tensor has the property of being always positive (including zero) since it is multiplied by the factor $(-1)^{N_{\rm t}}$ which is positive for an even number of transpositions   and negative for an odd number of transpositions, thus compensating for the negative sign coming from the Levi-Civita tensor.
The number of equivalent objects is indicated to the lower right of the terms. To see how this number is calculated see the caption in Fig.~(\ref{fig:cum}).

Let us study here the multivariate Gram-Charlier series expansion \citep[see][]{Berkowitz:1970:CMH}: 
\begin{equation}
P(\mbi\nu)=G(\mbi\nu)\left[1+ \sum^\infty_{l=1} \sum_{i_1\dots i_l}\frac{1}{l!}c_{i_1\dots {i_l}} (-1)^l h_{i_1\dots{i_l}}(\mbi\nu)\right]\,,
\end{equation} 
with $G(\mbi\nu)$ being a multivariate Gaussian distribution $G(\mbi\nu)=\frac{1}{\sqrt{2\pi}}e^{-\frac{\mbi\nu^\dagger\mbi\nu}{2}}$.

The corresponding multivariate Hermite polynomials are calculated by \citep[see][]{Berkowitz:1970:CMH}:  
  \begin{equation}
 h_{i_1\dots i_n}(\mbi \nu)=(-1)^n e^{\frac{1}{2}\mbi \nu^\dagger\mbi \nu}\frac{\partial^n}{\partial \nu_{i_1}\dots\partial \nu_{i_n}}e^{-\frac{1}{2}\mbi\nu^\dagger\mbi\nu} \, {,}
	\end{equation}
from which the following recursive formula can be built:
 \begin{equation}
 h_{i_1\dots i_n}(\mbi \nu)=(-1)^{n} e^{\frac{1}{2}\mbi \nu^\dagger\mbi \nu}\frac{\partial^n}{\partial \nu_{i_{n}}}(-1)^{n-1}e^{-\frac{1}{2}\mbi\nu^\dagger\mbi\nu} h_{i_1\dots i_{n-1}}(\mbi \nu) \, {,}
	\end{equation}
	which we have used in our calculations.

Here are the results for the first couple of polynomials: 
  \begin{eqnarray}
\label{eq:hemr3D}
		h_0(\mbi\nu)&=&1\\
  	h_i(\mbi\nu)&=&\nu_i \nonumber\\
  	h_{ij}(\mbi \nu)&=&\nu_i\nu_j- \delta^{\rm K}_{ij} \nonumber \\ 	
 		h_{ijk}(\mbi \nu)&=& \nu_i\nu_j\nu_k -\nu_i\delta^{\rm K}_{jk} -\nu_j\delta^{\rm K}_{ik}-\nu_k\delta^{\rm K}_{ij}\nonumber \\ 	
 		&& \hspace{-1.5cm}=\nu_i\nu_j\nu_k -\left[\frac{1}{2}\sum_{j_1j_2j_3\in[1,2,3]}\tilde{\epsilon}_{j_1j_2j_3}\nu_{i_{j_1}}\delta^{\rm K}_{i_{j_2}i_{j_3}}\right]_{3=\frac{3!}{2}}\nonumber \\ 	
 		h_{ijkl}(\mbi \nu)&=&\nu_i\nu_j\nu_k\nu_l\nonumber \\ 	
 		&&\hspace{-1.5cm}-\left[\frac{1}{2^2}\sum_{j_1\dots  j_4\in[1,\dots,4]}\tilde{\epsilon}_{j_1\dots  j_4}\nu_{i_{j_1}}\nu_{i_{j_2}}\delta^{\rm K}_{i_{j_3}i_{j_4}}\right]_{6=\frac{4!}{2^2}}\nonumber \\ 	
 		&&\hspace{-1.5cm}+\left[\frac{1}{2^3}\sum_{j_1\dots j_4\in[1,\dots,4]}\tilde{\epsilon}_{j_1\dots j_4}\delta^{\rm K}_{i_{j_1}i_{j_2}}\delta^{\rm K}_{i_{j_3}i_{j_4}}\right]_{3=\frac{4!}{2^3}} \nonumber \\ 
 		h_{ijklm}(\mbi \nu)&=&\nu_i\nu_j\nu_k\nu_l\nu_m\nonumber \\ 	
 		&&\hspace{-1.5cm}-\left[\frac{1}{3! 2}\sum_{j_1\dots j_5\in[1,\dots,5]}\tilde{\epsilon}_{j_1\dots j_5}\nu_{i_{j_1}}\nu_{i_{j_2}}\nu_{i_{j_3}}\delta^{\rm K}_{i_{j_4}i_{j_5}}\right]_{10=\frac{5!}{3!2}}\nonumber \\ 	
 		&&\hspace{-1.5cm}+\left[\frac{1}{2^3}\sum_{j_1\dots j_5\in[1,\dots,5]}\tilde{\epsilon}_{j_1\dots j_5}\nu_{i_{j_1}}\delta^{\rm K}_{i_{j_2}i_{j_3}}\delta^{\rm K}_{i_{j_4}i_{j_5}}\right]_{15=\frac{5!}{2^3}} \nonumber \\ 
 		h_{ijklmn}(\mbi \nu)&=&\nu_i\nu_j\nu_k\nu_l\nu_m\nu_n\nonumber \\ 	
 		&&\hspace{-1.5cm}-\left[\frac{1}{4! 2}\sum_{j_1\dots j_6\in[1,\dots,6]}\tilde{\epsilon}_{j_1\dots j_6}\nu_{i_{j_1}}\nu_{i_{j_2}}\nu_{i_{j_3}}\nu_{i_{j_4}}\delta^{\rm K}_{i_{j_5}i_{j_6}}\right]_{15=\frac{6!}{4! 2}}\nonumber \\ 	
 		&&\hspace{-1.5cm}+\left[\frac{1}{2^4}\sum_{j_1\dots j_6\in[1,\dots,6]}\tilde{\epsilon}_{j_1\dots j_6}\nu_{i_{j_1}}\nu_{i_{j_2}}\delta^{\rm K}_{i_{j_3}i_{j_4}}\delta^{\rm K}_{i_{j_5}i_{j_6}}\right]_{45=\frac{6!}{2^4}} \nonumber \\
 		&&\hspace{-1.5cm}-\left[\frac{1}{3! 2^3}\sum_{j_1\dots j_6\in[1,\dots,6]}\tilde{\epsilon}_{j_1\dots j_6}\delta^{\rm K}_{i_{j_1}i_{j_2}}\delta^{\rm K}_{i_{j_3}i_{j_4}}\delta^{\rm K}_{i_{j_5}i_{j_6}}\right]_{15=\frac{6!}{3! 2^3}}  \nonumber \, {,}
	\end{eqnarray}
	where we have used the same notation as for the moments. One should note that the number of equivalent terms for the moments and Hermite polynomials coincides with the factors in the corresponding terms of the univariate case.
The Gram-Charlier coefficients can now be calculated by making an ensemble average over the Hermite polynomials: 
\begin{equation}
c_{i_{1}\dots i_{l}}=(-1)^l\int \dd \mbi\nu P(\mbi\nu)\,h_{i_{1}\dots i_{l}}(\mbi\nu)=(-1)^l \langle h_{i_{1}\dots i_{l}}(\mbi\nu)\rangle\,.
\end{equation} 
The first coefficient is $c_0=1$ as in the univariate case and the rest is calculated using the above equation:  
\begin{eqnarray}
\label{eq:gramMV}
c_i&=&0\\
c_{ij}&=&0\nonumber\\
c_{ijk}&=&-\mu_{ijk}=-\kappa_{ijk}\nonumber\\
c_{ijkl}&=&\mu_{ijkl}=\kappa_{ijkl}\nonumber\\
c_{ijklm}&=&-\mu_{ijklm}\nonumber\\
&&\hspace{-1.5cm}-\left[\frac{1}{3! 2}\sum_{j_1\dots j_5\in[1,\dots,5]}\tilde{\epsilon}_{j_1\dots j_5}\mu_{i_{j_1}i_{j_2}i_{j_3}}\delta^{\rm K}_{i_{j_4}i_{j_5}}\right]_{10}\nonumber\\
&=&-\kappa_{ijklm}\nonumber\\
c_{ijklmn}&=&\mu_{ijklmn}\nonumber\\
&&\hspace{-1.5cm}-\left[\frac{1}{4! 2}\sum_{j_1\dots j_6\in[1,\dots,6]}\tilde{\epsilon}_{j_1\dots j_6}\mu_{i_{j_1}i_{j_2}i_{j_3}i_{j_4}}\delta^{\rm K}_{i_{j_5}i_{j_6}}\right]_{15}\nonumber \\
 		&&\hspace{-1.5cm}+\left[\left(\frac{1}{2^4}-\frac{1}{3! 2^3}\right)\sum_{j_1\dots j_6\in[1,\dots,6]}\tilde{\epsilon}_{j_1\dots j_6}\delta^{\rm K}_{i_{j_1}i_{j_2}}\delta^{\rm K}_{i_{j_3}i_{j_4}}\delta^{\rm K}_{i_{j_5}i_{j_6}}\right]_{30} \nonumber\\
 		 	&=&\kappa_{ijklmn}\nonumber\\
 		&&\hspace{-1.5cm}+\left[\frac{1}{3!3!2}\sum_{j_1\dots j_6\in[1,\dots,6]}\tilde{\epsilon}_{j_1\dots j_6}\kappa_{i_{j_1}i_{j_2}i_{j_3}}\kappa_{i_{j_4}i_{j_5}i_{j_6}}\right]_{10} \nonumber\,.
\end{eqnarray}
Rearranging terms in the Gram-Charlier series based on the findings in the univariate case we can build the multivariate Edgeworth expansion: 
 \begin{eqnarray}	
\label{eq:edge1}
\lefteqn{P(\mbi\Phi)= \frac{\dd \mbi \nu}{\dd\mbi\Phi} P(\mbi\nu)=({\rm det }(\mat S))^{-1/2}G(\mbi\nu)}\\
&&\hspace{-1.cm}\times\left[1+\frac{1}{3!}\sum_{ijk}\langle\nu_{i}\nu_{j}\nu_{k}\rangle_{\rm c} h_{ijk}(\mbi\nu)+\frac{1}{4!}\sum_{ijkl}\langle\nu_{i}\nu_{j}\nu_{k}\nu_{l}\rangle_{\rm c} h_{ijkl}(\mbi\nu)\right.\nonumber\\
&&\hspace{-1.cm}+\frac{1}{6!}\sum_{ijklmn}\left[\frac{1}{3!3!2}\sum_{j_1\dots j_6\in[1,\dots,6]}\right.\nonumber\\
&&\hspace{-1.cm}\left.\left.\times\tilde{\epsilon}_{j_1\dots j_6}\langle\nu_{i_{j_1}}\nu_{i_{j_2}}\nu_{i_{j_3}}\rangle_{\rm c}\langle\nu_{i_{j_4}}\nu_{i_{j_5}}\nu_{i_{j_6}}\rangle_{\rm c}\right]_{10}h_{ijklmn}(\mbi\nu)+\dots\right]\nonumber\,,	
\end{eqnarray} 
which can also be written as:
\begin{eqnarray}
\label{eq:edge}
\lefteqn{P(\mbi\Phi)= ({\rm det }(\mat S))^{-1/2}G(\mbi\nu)}\\
&&\hspace{-1.cm}\times\left[1+\frac{1}{3!}\sum_{i'j'k'}\langle\Phi_{i'}\Phi_{j'}\Phi_{k'}\rangle_{\rm c}\sum_{ijk}S_{ii'}^{-1/2}S_{jj'}^{-1/2}S_{kk'}^{-1/2} h_{ijk}(\mbi\nu)\right.\nonumber\\
&&\hspace{-1.cm}+\frac{1}{4!}\sum_{i'j'k'l'}\langle\Phi_{i'}\Phi_{j'}\Phi_{k'}\Phi_{l'}\rangle_{\rm c}\sum_{ijkl}S_{ii'}^{-1/2}S_{jj'}^{-1/2}S_{kk'}^{-1/2}S_{ll'}^{-1/2} h_{ijkl}(\mbi\nu)\nonumber\\
&&\hspace{-1.cm}+\frac{1}{6!}\sum_{i'j'k'l'm'n'}\nonumber\\
&&\hspace{-1.cm}\times\left[\frac{1}{3!3!2}\sum_{j_1\dots j_6\in[1,\dots,6]}\tilde{\epsilon}_{j_1\dots j_6}\langle\Phi_{i'_{j_1}}\Phi_{i'_{j_2}}\Phi_{i'_{j_3}}\rangle_{\rm c}\langle\Phi_{i'_{j_4}}\Phi_{i'_{j_5}}\Phi_{i'_{j_6}}\rangle_{\rm c}\right]_{10}\nonumber\\
&&\hspace{-1.cm}\times\left.\sum_{ijklmn}S_{ii'}^{-1/2}S_{jj'}^{-1/2}S_{kk'}^{-1/2}S_{ll'}^{-1/2}S_{mm'}^{-1/2}S_{nn'}^{-1/2} h_{ijklmn}(\mbi\nu)+\dots\right]\nonumber\,,
\end{eqnarray} 
where we have inserted the expression for the field $\mbi \nu$ and introduced the following notation: $i'\equiv i'_1,j'\equiv i'_2,k'\equiv i'_3,l'\equiv i'_4,m'\equiv i'_5,n'\equiv i'_6$.
 From this expression it is possible to calculate the probability of a density field $\mbi\Phi$ 
 given the higher order point correlation functions (of the logarithm of the density field!).  

We can find a more compact expression for the last equation if we define a skewness term accounting for the asymmetry of the distribution function as:
\begin{equation}
\mathcal S (\mbi\nu)\equiv \frac{1}{3!}\sum_{ijk}\kappa_{ijk}{h}_{ijk}(\mbi \nu)\,,
\end{equation}
and a kurtosis term accounting for the flatness of the distribution function  composed by two terms: $\mathcal K\equiv\mathcal K_{\rm A}+\mathcal K_{\rm B}$
with the first term given by:
\begin{equation}
\mathcal K_{\rm A} (\mbi\nu)\equiv \frac{1}{4!}\sum_{ijkl}\kappa_{ijkl}{h}_{ijkl}(\mbi \nu)\,,
\end{equation}
and the second term given by:
\begin{eqnarray}
\lefteqn{\mathcal K_{\rm B} (\mbi\nu)\equiv} \\
&&\hspace{-1.cm}\frac{1}{6!}\sum_{ijklmn}\left[\frac{1}{3!3!2}\sum_{j_1\dots j_6\in[1,\dots,6]}\tilde{\epsilon}_{j_1\dots j_6}\kappa_{i_{j_1}i_{j_2}i_{j_3}}\kappa_{i_{j_4}i_{j_5}i_{j_6}}\right]_{10}h_{ijklmn}(\mbi\nu)\nonumber\,.
\end{eqnarray}

Now we can rephrase Eq.~(\ref{eq:edge}) by:
\begin{eqnarray}
\label{eq:skewedneg}
\lefteqn{P(\mbi \nu)=G(\mbi\nu)\left[1+\mathcal S(\mbi\nu)+\mathcal K (\mbi\nu)+\dots\right]}\\
&&=G(\mbi\nu)\left[1+\frac{1}{6}\sum_{ijk}\kappa_{ijk}{h}_{ijk}(\mbi \nu)+\frac{1}{24}\sum_{ijkl}\kappa_{ijkl}{h}_{ijkl}(\mbi \nu)\right.\nonumber\\
 &&\hspace{-1.2cm}\left.+\frac{1}{720}\sum_{ijklmn}\left[\frac{1}{72}\sum_{j_1\dots j_6\in[1,\dots,6]}\tilde{\epsilon}_{j_1\dots j_6}\kappa_{i_{j_1}i_{j_2}i_{j_3}}\kappa_{i_{j_4}i_{j_5}i_{j_6}}\right]_{10}h_{ijklmn}(\mbi\nu)\right.\nonumber\\
 &&\hspace{-1.2cm}\left.+\dots\right]\nonumber\,.
\end{eqnarray}

Note that the last term in the previous equation has 10 members (6 indices: $6!/72$) which will only be equal under certain symmetry conditions. Our calculations confirm that the first three terms in Eq.~(\ref{eq:skewedneg}) are a trivial generalization of the univariate case (see Eq.~\ref{eq:edge1V}), whereas the 4th term is not.

\subsubsection{On the positive definiteness of the expanded Normal distribution}

 Please note that $P_k(\mbi \nu)$ is not a real distribution function as it is not generally positive definite. The truncated expanded distribution function should be regarded only as an approximation at $k$-th order (1st order: lognormal, 2nd order: includes skewness $\mathcal S$, 3rd order: includes skewness $\mathcal S$ and kurtosis $\mathcal K$). To ensure a positive definiteness one has to impose  the additional condition: $1+\mathcal S(\mbi  \nu)\ge0,\, \forall \mbi \nu$ at 2nd order, $1+\mathcal S(\mbi  \nu)+\mathcal K(\mbi  \nu)\ge0,\, \forall \mbi \nu$ at 3rd order, etc..
It is also possible to make an exponenzation of the non-Gaussian contribution or a quadratic expression  \citep[see the works on non-Gaussian realizations in the CMB:][]{2000ApJ...534...25C,2001PhRvD..63j3512C}: 
\begin{equation}
\label{eq:skewedpos}
P_k(\mbi \nu)=G(\mbi\nu)\exp\left(\mathcal S(\mbi\nu)+\mathcal K (\mbi\nu)+\dots\right)\,.
\end{equation}
For small skewness and kurtosis terms the Taylor expansion of the exponential will be dominated by $1+\mathcal S(\mbi\nu)+\mathcal K (\mbi\nu)+\dots$, so that Eq.~(\ref{eq:skewedpos}) will be very close to Eq.~(\ref{eq:skewedneg}) ensuring positive definiteness.
In general one has to define a function $\mathcal F$ of the non-Gaussian contributions which ensures positive definiteness: 
\begin{equation}
\label{eq:skewedpos2}
P_k(\mbi \nu)=G(\mbi\nu)\mathcal F\left(1+\mathcal S(\mbi\nu)+\mathcal K (\mbi\nu)+\dots\right)\,.
\end{equation}

\begin{figure}
  \hspace{1.cm}
\setlength{\unitlength}{0.254mm}
\begin{picture}(264,161)(115,-180)
        \special{color rgb 0 0 0}\put(125,-76){\shortstack{linear}} 
        \special{color rgb 0 0 0}\put(120,-116){\shortstack{regime}} 
        \special{color rgb 0 0 0}\put(115,-156){\shortstack{nonlinear}} 
        \special{color rgb 0 0 0}\put(215,-66){\shortstack{Gaussian}} 
        \special{color rgb 0 0 0}\put(195,-91){\shortstack{expanded Gaussian}} 
        \special{color rgb 0 0 0}\put(215,-116){\shortstack{Lognormal}} 
        \special{color rgb 0 0 0}\put(190,-141){\shortstack{expanded Lognormal}} 
        \special{color rgb 0 0 0}\put(180,-31){\shortstack{Hierarchy of matter statistics}} 
        \special{color rgb 0 0 0}\put(210,-166){\shortstack{full nonlinear}} 
        \special{color rgb 0 0 0}\allinethickness{0.254mm}\put(165,-80){\vector(0,-1){100}} 
        \special{color rgb 0 0 0}\allinethickness{0.254mm}\put(165,-140){\vector(0,1){95}} 
        \special{color rgb 0 0 0}\allinethickness{0.254mm}\put(355,-45){\vector(0,-1){135}} 
        \special{color rgb 0 0 0}\allinethickness{0.254mm}\put(320,-180){\vector(0,1){135}} 
        \special{color rgb 0 0 0}\put(325,-76){\shortstack{scale}} 
        \special{color rgb 0 0 0}\put(360,-156){\shortstack{time}} 
        \special{color rgb 0 0 0} 
\end{picture}
\caption{The matter statistics depends on the gravitational regime, the cosmic scale and the cosmic time we are looking at. Towards large scales, in the linear regime matter is closely Gaussian distributed. At smaller scales and later times gravitational clustering will start to depart from Gaussianity. When looking at small deviations from Gaussianity a higher order correlation expansion can be done. The lognormal distribution is a good description for the further nonlinear regime as it can be regarded as linear in a Lagrangian framework which is known to be valid in the quasi-nonlinear regime. At even lower scales and late times the lognormal distribution fails and an expansion around this distribution function can be done. When {\it shell-crossing}  starts and structures form caustics (the full nonlinear regime) the statistical state-of-the-art description is based on numerical N-body simulations.  }
\label{fig:hier}
\end{figure}
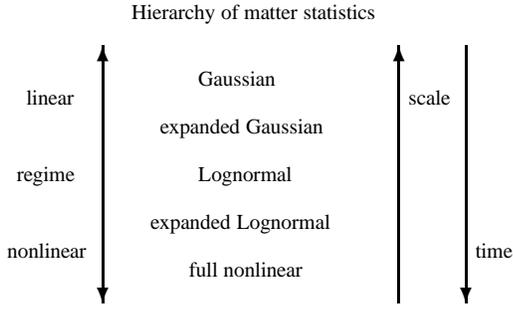

\subsubsection{Lognormal limit}

\label{sec:loglim}

On scales larger than about 10 Mpc the lognormal distribution resembles very well the observed galaxy and matter density statistics \citep[see][]{1934ApJ....79....8H,2005MNRAS.356..247W,kitaura_sdss}. 
It was demonstrated by \citet[][]{kitaura_log} that the lognormal prior can be applied at least down to scales of few Mpc to fit the matter statistics in the overdense regions.   
Towards the linear regime in the mild non-linear regime the correction terms in the Edgeworth expansion start to become negligible and we can model the density field with a multivariate lognormal distribution \citep[see][]{1991MNRAS.248....1C}:
\begin{eqnarray}
\label{eq:LOGNORMAL}
\lefteqn{{ P}(\mbi\delta_{{\rm M}}|\mat S)=\frac{1}{\sqrt{(2\pi)^{N_{\rm c}} {\rm det} (\mat S)}}\prod_k \frac{1}{1+\delta_{{\rm M}k}}} \\
&&\hspace{-0.5cm}\times {\rm exp}\left[{-\frac{1}{2}\sum_{ij} \left({\rm ln}(1+\delta_{{\rm M}i})-\mu_{si}\right) S^{-1}_{ij} \left({\rm ln}(1+\delta_{{\rm M}j})-\mu_{sj}\right)}\right]\nonumber \, {,}
\end{eqnarray}
where $\mat S$ is the covariance matrix of the lognormal distribution.
Note that the covariance matrix is defined by:  $S_{ij}\equiv\langle s_is_j\rangle-\mu_{si}\mu_{sj}$ (which does not coincide with the covariance of the overdensity field: $\langle\delta_{{\rm M}i}\delta_{{\rm M}j}\rangle$)  and \(\mu_{si}\) describes a constant mean field given by (a derivation of both the covariance and the mean can be found in appendix \ref{app:multivar}):
\begin{equation}
\label{eq:MU}
\mu_{si}=-\frac{1}{2}\sum_{ij} \hat{\hat{S}}_{ij}=-\frac{1}{2}S_{ii} \, ,
\end{equation}
with the hats denoting the Fourier transform of the auto-correlation matrix. The constant term $S_{ii}$ corresponds to the correlation function evaluated at zero: $S[0]$, i.e.~when it is evaluated for the distance of an object at a certain position with itself.

\subsubsection{Gaussian limit}
\label{sec:gauss}

On very large scales ($\gsim1 100$ Mpc), in the linear regime, the information about the primordial fluctuations has been preserved throughout cosmic history. Inflationary scenarios predict a close to Gaussian distribution function for the initial density fluctuations \citep[see][]{1981PhRvD..23..347G, 1982PhRvL..49.1110G, 1982PhLB..117..175S, 1982CMaPh..87..395H,  1982PhLB..108..389L, 1982PhRvL..48.1220A, 1983PhRvD..28..679B}. Therefore we expect to obtain the Gaussian prior in the low density limit.  
When $|\delta_{\rm M}|\ll 1$ the signal $s$ can be approximated by the linear term of the Mercator series ${\rm ln}(1+\delta_{{\rm M}j})\sim\delta_{{\rm M}j}$ and the mean is approximately zero $\mu_{si}\sim0$. 
In that case the prior distribution for the matter density field is given by the Gaussian multivariate distribution function \citep[see][]{1986ApJ...304...15B}:
\begin{eqnarray}
\label{eq:GAUSS}
\lefteqn{{ P}(\mbi \delta_{{\rm M}}|\mat S)=} \\
&&\frac{1}{\sqrt{(2\pi)^{N_{\rm c}} {\rm det} (\mat S)}}{\rm exp}\left({-\frac{1}{2}\sum_{ij} \delta_{{\rm M}i} S^{-1}_{ij} \delta_{{\rm M}j}}\right) \, {.}\nonumber
\end{eqnarray}
Please note that the correlation matrix $\mat S$ is now defined by: $S_{ij}\equiv\langle \delta_{{\rm M}i}\delta_{{\rm M}j}\rangle$.
This prior is the one required by the Bayesian approach to derive the Wiener-filter \citep[see][]{zaroubi} and has been extensively applied to CMB-mapping \citep[see for example][]{1994ApJ...432L..75B,1997ApJ...480L..87T} and to the large-scale structure reconstruction \citep[see for example][]{zaroubi,2004MNRAS.352..939E,2006MNRAS.373...45E,kitaura_sdss}.

Intrinsic deviations from the Gaussian or the mild nonlinear gravitational regime may be described by expanding the Gaussian distribution in an Edgeworth expansion.
Note that the whole multivariate formalism presented in this section can also be  applied to study weakly non-Gaussian matter fields as was shown by  \citet[][]{1995ApJ...442...39J} for the univariate case. One has to define instead $\mbi\nu\equiv \mat S^{-1/2}\mbi \delta_{{\rm M}}$ and use the higher order correlations corresponding to the overdensity field. However, additional problems could arise here yielding negative densities which are not present in the lognormal formulation. 

The matter distributions described by the Eqs.~(\ref{eq:LOGNORMAL}) and (\ref{eq:GAUSS})  are only valid in a limited range of overdensities.
In order to be able to go further into the nonlinear regime we need to model the higher order statistics.
The problem we face here is that we have to introduce more complex models increasing the number of parameters \citep[see the works by][including galaxy biasing and redshift distortions in higher order correlations]{1998ApJ...496..586S,1999ApJ...522...46T,2003ApJ...584....1M,2004ApJ...614..527Z,2008PhRvD..78h3519M}.

The simplest non-trivial higher order correlation representation is given by the hierarchical models which we will discuss in the next section.

\renewcommand{\arraystretch}{1.5}

\begin{table*}
\begin{tabular}{cc|c|c|c|c|c|c|l}
\multicolumn{8}{|c|}{Combinatorics of higher order correlation functions in the hierarchical model} \\\cline{1-8}
\multicolumn{1}{|c|}{$n$: order of the correlation function}& & 3&  & 4 &  & 5 & \\\cline{1-8}
\multicolumn{1}{|c|}{$m=2(n-1)$: number of indices} & & 4 &  & 6 &  & 8 & \\\cline{1-8}
\multicolumn{1}{|c|}{$N_n^\alpha$: number of trees} & & 1 &  & 2 &  &3 & \\\cline{1-8}
\multicolumn{1}{|c|}{$N^\alpha_{n\beta}\equiv\frac{n!}{N^\alpha_{n\gamma}}$: number of leafs}& & \hspace{-0.1cm}$(i|j|k)$\hspace{0.1cm} & $N^\alpha_{3\beta}$ & \hspace{-0.1cm}$(i|j|k|l)$\hspace{0.1cm} & $N^\alpha_{4\beta}$ & \hspace{-0.1cm}$(i|j|k|l|m)$\hspace{0.1cm}  & $N^\alpha_{5\beta}$ \\ \cline{1-8}
\multicolumn{1}{|c|}{\multirow{3}{*}{$\alpha$: tree}} &
\multicolumn{1}{|c|}{1}  & $(2|1|1)\big|_3$& $3=\frac{3!}{2}$ & $(3|1|1|1)\big|_4$ & $4=\frac{4!}{3!}\frac{2}{2}$ & $(4|1|1|1|1)\big|_5$ &$5=\frac{5!}{4!}\frac{3!}{3!}$   \\ \cline{2-8}
&\multicolumn{1}{|c|}{2} &          &  & $(2|2|1|1)\big|_{12}$ & $12=\frac{4!}{2^2}2$ & $(3|2|1|1|1)\big|_{60}$ & $60=\frac{5!}{3!}\frac{3!}{2}$  \\ \cline{2-8}
&\multicolumn{1}{|c|}{3} &          &  &             &  & $(2|2|2|1|1)\big|_{60}$ & $60=\frac{5!}{3!2}3!$    \\ \cline{1-8}
\multicolumn{3}{|c|}{leafs of the three-point correlation function}&\multicolumn{5}{|c|}{$(2|1|1)\big|_3$: $(\underbrace{\xi_{ij}\xi_{ik}}_{(2|1|1)},\underbrace{\xi_{ij}\xi_{jk}}_{(1|2|1)},\underbrace{\xi_{ik}\xi_{jk}}_{(1|1|2)})$} \\ \cline{1-8}
\multicolumn{3}{|c|}{leafs of the first tree of the four-point correlation function}&\multicolumn{5}{|c|}{$(3|1|1|1)\big|_4$: $(\underbrace{\xi_{ij}\xi_{ik}\xi_{il}}_{(3|1|1|1)},\underbrace{\xi_{ij}\xi_{jk}\xi_{jl}}_{(1|3|1|1)},\underbrace{\xi_{ik}\xi_{jk}\xi_{kl}}_{(1|1|3|1)},\underbrace{\xi_{il}\xi_{jl}\xi_{kl}}_{(1|1|1|3)})$} \\ \cline{1-8}
\multicolumn{8}{|c|}{leafs of the second tree of the four-point correlation function} \\
\multicolumn{8}{|c|}{$(2|2|1|1)\big|_{12}$: $(\underbrace{\xi_{ij}\xi_{ik}\xi_{jl},\xi_{ij}\xi_{il}\xi_{jk}}_{(2|2|1|1)},\underbrace{\xi_{ij}\xi_{ik}\xi_{kl},\xi_{il}\xi_{ik}\xi_{kj}}_{(2|1|2|1)},\underbrace{\xi_{ij}\xi_{il}\xi_{kl},\xi_{ik}\xi_{il}\xi_{jl}}_{(2|1|1|2)},\underbrace{\xi_{ij}\xi_{jk}\xi_{kl},\xi_{ik}\xi_{jk}\xi_{jl}}_{(1|2|2|1)},\underbrace{\xi_{ij}\xi_{jl}\xi_{kl},\xi_{il}\xi_{jk}\xi_{jl}}_{(1|2|1|2)},\underbrace{\xi_{ik}\xi_{jl}\xi_{kl},\xi_{il}\xi_{jk}\xi_{kl}}_{(1|1|2|2)})$} \\ \cline{1-8}
\end{tabular}
\caption{The different trees up to order 5 are shown with the corresponding number of leafs/labellings. The trees are constructed by distributing the $m$ indices  $n$ times so that there is at least one index for each dimension of $n$ neglecting the index order. Note, that by the same procedure one can find that the sixth order correlation function has 5 trees instead of 4 as one would naivly expect. To know then the number of leafs we have not only to consider the index order, but also the couple combinations of indices. The first factor can be calculated in an analogous way to the number of elements in Fig.~(\ref{fig:cum}). To calculate the latter factor one has to first subtract $n$ indices to  the tree (since we are interested in the couple combinations we already assume that each correlation function has been assigned one index)  and then divide $(m-n)!$ by the factorials of  the remaining numbers assigned to each index. For example for the second tree of the fifth order we would get after subtraction: $(2|1|0|0|0)$, which has three indices left (3! permutations) with 2 equal ones (divided by 2): $3!/2$. Dividing this number by $3!$ which comes from the three indices with equal number of appearances, we get that the leafs redundancy number is: $N^2_{5\gamma}=2$.
The leafs for the three-point and four-point correlation function are shown in detail at the bottom of the table. } 
\label{fig:comb}
\end{table*}

\subsection{Hierarchical model for higher order correlation functions}
\label{sec:HM}

Hierarchical models rely on the assumption that the cosmological structures are to some extent self-similar. 
In these models higher order correlation functions are constructed from products of the two-point correlation function
\citep[see][]{1978ApJ...221...19F,1984ApJ...277L...5F,1986ApJ...306..358F,1989A&A...220....1B}:
\begin{equation}
\label{eq:HM}
\xi_{i_1\dots i_n}\equiv \langle\Phi_{i_1}\dots \Phi_{i_n}\rangle_\cc= \sum_\alpha Q^\alpha_n\sum_{\mathcal L_\alpha}\prod^{n-1} \xi_{ij}  \, ,
\end{equation}
such that the whole set of points $i_1\dots i_n$ is connected by links of $\xi_{ij}$.
These links are organized in a tree structure, where $\alpha$ are the trees corresponding to each order $n$. The sum over $\mathcal L_\alpha$ denotes a sum over all possible labellings or leafs of a given tree.  The remaining freedom is encoded in the $Q^\alpha_n$ hierarchical coefficients for each tree $\alpha$ and order $n$.
 Note that the hierarchical model always refers to the overdensity field $\mbi\delta_{\rm M}$, but we are assuming that it also  applies for the field $\mbi\Phi$ defined in Eq.~(\ref{eq:phi}). 

Particular expressions for the three and four-point correlation functions were already proposed by \citet[see][]{1978ApJ...221...19F}.
The three-point correlation function can then be written as:
\begin{equation}
\label{eq:HM3}
\xi_{i_1\dots i_3} 
= Q_3\left[\frac{1}{2}\sum_{j_1j_2 j_3\in[1,2,3]}\tilde{\epsilon}_{j_1j_2j_3}\xi_{i_{j_1}i_{j_2}}\xi_{i_{j_1}i_{j_3}} \right]_{3} \, ,
\end{equation}
where we have denoted the only one hierarchical coefficient for the three order case as $Q_3$ (we show in Tab.~\ref{fig:comb} how to calculate the number of trees and the number of corresponding leafs). 
Note that the three-point correlation function is the sum of the leafs for the single tree:
\begin{equation}
\label{eq:HM32}
\xi_{ijk}= Q_3\left[\xi_{ij}\xi_{ik}+\xi_{ij}\xi_{jk}+\xi_{ik}\xi_{jk}\right] \, .
\end{equation}
 
The four-point correlation function has 16 leafs, 4 leafs in the first tree and 12 leafs in the second one \citep[see][and our calculation in Tab.~\ref{fig:comb}]{1978ApJ...221...19F,1989A&A...220....1B}:
\begin{eqnarray}
\label{eq:HM4}
\lefteqn{\xi_{i_1\dots i_4}} \nonumber\\
&& =Q_4^{\rm a}\left[\frac{1}{3!}\sum_{j_1\dots j_4\in[1,\dots,4]}\tilde{\epsilon}_{j_1\dots j_4}\xi_{i_{j_1}i_{j_2}}\xi_{i_{j_1}i_{j_3}}\xi_{i_{j_1}i_{j_4}}\right]_{4}\nonumber\\
&&+ Q_4^{\rm b}\left[\frac{1}{2}\sum_{j_1\dots j_4\in[1,\dots,4]}\tilde{\epsilon}_{j_1\dots j_4}\xi_{i_{j_1}i_{j_2}}\xi_{i_{j_2}i_{j_3}}\xi_{i_{j_3}i_{j_4}}  \right]_{12}\,,
\end{eqnarray}
which can also be  written as:
\begin{eqnarray}
\label{eq:HM42}
\lefteqn{\xi_{ijkl}=}\nonumber\\
&&Q_4^{\rm a}\left[ \xi_{ij}\xi_{ik}\xi_{il}+\xi_{ij}\xi_{jk}\xi_{jl}+\xi_{ik}\xi_{jk}\xi_{kl}+\xi_{il}\xi_{jl}\xi_{kl} \right]\nonumber\\
&&+Q_4^{\rm b}\left[\xi_{ij}\xi_{ik}\xi_{jl}+\xi_{ij}\xi_{il}\xi_{jk}+\xi_{ij}\xi_{ik}\xi_{kl}+\xi_{il}\xi_{ik}\xi_{kj}  \right.\nonumber\\
&&\left.+ \xi_{ij}\xi_{il}\xi_{kl}+\xi_{ik}\xi_{il}\xi_{jl}+\xi_{ij}\xi_{jk}\xi_{kl}+\xi_{ik}\xi_{jk}\xi_{jl}\right.\nonumber\\
&&\left.+ \xi_{ij}\xi_{jl}\xi_{kl}+\xi_{il}\xi_{jk}\xi_{jl}+\xi_{ik}\xi_{jl}\xi_{kl}+\xi_{il}\xi_{jk}\xi_{kl} \right]\, .
\end{eqnarray}
where we have denoted the first hierarchical coefficient as $Q_4^{\rm a}$ and the second one as  $Q_4^{\rm b}$.

\subsection{Non-Gaussian multivariate Edgeworth expansion with the hierarchical model}
\label{sec:edgeHM}

We can use now the three and four-point correlation functions calculated from the hierarchical model to apply the third order Edgeworth expansion to the lognormal probability distribution. Note that the results of this section can also be applied for the Gaussian case by inserting the correlation functions of the overdensity field instead of the correlation functions of the logarithm of the density field. 
In the next subsections we present the skewness and kurtosis terms which are required in the Edgeworth expansion (see Eq.~\ref{eq:edge}).

\subsubsection{Skewness terms}

The second order term in the Edgeworth expansion describes the skewness with respect to the lognormal distribution and requires the third order Hermite polynomial (see Eq.~\ref{eq:hemr3D}).
Let us define the weighted Hermite polynomial:
\begin{eqnarray}
\label{eq:herm}
\tilde{h}_{i'j'k'}(\mbi \nu)&\equiv& \sum_{ijk}S_{ii'}^{-1/2}S_{jj'}^{-1/2}S_{kk'}^{-1/2}  h_{ijk}(\mbi \nu) \\
&=& \eta_{i'}\eta_{j'}\eta_{k'} -\eta_{i'}S_{j'k'}^{-1}-\eta_{j'}S_{i'k'}^{-1}-\eta_{k'}S_{i'j'}^{-1} \nonumber\,,
\end{eqnarray}
with 
\begin{equation}
\eta_{i}\equiv\sum_j S_{ij}^{-1} \Phi_j\,.
\end{equation}
Then we can calculate the skewness term $\mathcal S$ as (see Eq.~\ref{eq:edge} and appendix \ref{app:skew}): 
\begin{eqnarray}
\label{eq:skew} 
\lefteqn{\mathcal S (\mbi\nu)\equiv \frac{1}{3!}\sum_{ijk}\kappa_{ijk}{h}_{ijk}(\mbi \nu)=\frac{1}{3!}\sum_{i'j'k'}\xi_{i'j'k'}\tilde{h}_{i'j'k'}(\mbi \nu)}\\
\hspace{-1cm}&&=\frac{Q_3}{3!}\sum_{i'j'k'}\left[S_{i'j'}S_{i'k'}+S_{i'j'}S_{j'k'}+S_{i'k'}S_{j'k'}\right]\tilde{h}_{i'j'k'}(\mbi \nu)\nonumber\,\\
\hspace{-1.cm}&&=Q_3\left[\frac{1}{2}\sum_{i}\Phi_{i}^2\eta_{i}-\frac{1}{2}\sum_{i}S_{ii}\eta_{i}-\sum_{i}\Phi_{i}\right]\nonumber\,,
\end{eqnarray}
where we have identified the two-point correlation function  $\xi_{ij}$ with $S_{ij}$.

\subsubsection{Kurtosis terms}

The third order term in the Edgeworth expansion requires the fourth and sixth order Hermite polynomials and the three and four-point correlation functions.
Thus, we separate the kurtosis $\mathcal K$ into two contributions $\mathcal K\equiv\mathcal K_{\rm A}+\mathcal K_{\rm B}$.
	 
Let us start with the first contribution which requires the fourth order Hermite polynomial:
\begin{eqnarray}
\label{eq:hermkurt}
\tilde{h}_{i'j'k'l'}(\mbi \nu)&\equiv& \sum_{ijk}S_{ii'}^{-1/2}S_{jj'}^{-1/2}S_{kk'}^{-1/2}S_{ll'}^{-1/2}  h_{ijkl}(\mbi \nu) \\
&=& \eta_{i'}\eta_{j'}\eta_{k'}\eta_{l'} \nonumber\\ 
&&\hspace{-.5cm}-\left[\frac{1}{2^2}\sum_{j_1\dots  j_4\in[1,\dots,4]}\tilde{\epsilon}_{j_1\dots  j_4}\eta_{i'_{j_1}}\eta_{i'_{j_2}}S^{-1}_{i'_{j_3}i'_{j_4}}\right]_{6}\nonumber \\ 	
&&\hspace{-.5cm}+\left[\frac{1}{2^3}\sum_{j_1\dots j_4\in[1,\dots,4]}\tilde{\epsilon}_{j_1\dots j_4}S^{-1}_{i'_{j_1}i'_{j_2}}S^{-1}_{i'_{j_3}i'_{j_4}}\right]_{3}\nonumber\,.
\end{eqnarray}
The first contribution $\mathcal K_{\rm A}$  to the kurtosis term $\mathcal K$ can be written as (see Eq.~\ref{eq:edge}): 
\begin{eqnarray}
\label{eq:kurt1} 
\hspace{-.5cm}\mathcal K_{\rm A} (\mbi\nu)&\equiv& \frac{1}{4!}\sum_{ijkl}\kappa_{ijkl}{h}_{ijkl}(\mbi \nu)=\frac{1}{4!}\sum_{i'j'k'l'}\xi_{i'j'k'l'}\tilde{h}_{i'j'k'l'}(\mbi \nu)
\nonumber\,.
\end{eqnarray}
which after some calculations (see appendix \ref{app:kurt1}) leads to:
\begin{eqnarray}
\label{eq:kurt12} 
\lefteqn{\mathcal K_{\rm A} (\mbi\nu)=} \\
&&\hspace{-0.5cm}\frac{Q_4^{\rm a}}{2}\left[ \frac{1}{3}\sum_{i}\Phi_{i}^3\eta_i-\sum_{i}\Phi^2_{i}\right]\nonumber\\
&&\hspace{-0.5cm} -\left(\frac{Q_4^{\rm a}}{2}+Q_4^{\rm b}\right) \sum_{i}S_{ii}\Phi_{i}\eta_{i}  +\frac{1}{2}\left(Q_4^{\rm a}+Q_4^{\rm b}\right)\sum_{i}S_{ii}\nonumber\\
&&\hspace{-0.5cm} +\frac{Q_4^{\rm b}}{2}\left[\sum_{ij}\eta_{i}\Phi_{i}S_{ij}\Phi_{j}\eta_{j}-\sum_{ij}\eta_{i}S^2_{ij}\eta_{j}\right.\nonumber\\
&&\hspace{-0.5cm} \left.-\left(\sum_{i}\Phi_{i}\right)^2-{2}\sum_{ij}\Phi_{i}S_{ij}\eta_{j}+2\sum_{ij}S_{ij}\right] \nonumber\,.
\end{eqnarray}

As we saw in sections (\ref{sec:unicase}) and (\ref{sec:MV}) the asymptotic Edgeworth expansion has a second term at third order. It is in particular the fifth term of the sixth-order moment in Fig.~(\ref{fig:cum}) which is the product of two three-point correlation functions.
Accordingly, the second contribution $\mathcal K_{\rm B}$ requires the sixth order weighted Hermite polynomial:
\begin{eqnarray}
\label{eq:herm6}
\lefteqn{\tilde{h}_{i'j'k'l'm'n'}(\mbi \nu)}\\
&&\equiv \sum_{ijk}S_{ii'}^{-1/2}S_{jj'}^{-1/2}S_{kk'}^{-1/2}S_{ll'}^{-1/2}S_{mm'}^{-1/2}S_{nn'}^{-1/2}  h_{ijklmn}(\mbi \nu) \nonumber\\
&&= \eta_{i'}\eta_{j'}\eta_{k'} \eta_{l'}\eta_{m'}\eta_{n'}\nonumber \\ 	
&&\hspace{0cm}-\left[\frac{1}{4! 2}\sum_{j_1\dots j_6\in[1,\dots,6]}\tilde{\epsilon}_{j_1\dots j_6} \eta_{i'_{j_1}} \eta_{i'_{j_2}}\eta_{i'_{j_3}}\eta_{i'_{j_4}} S^{-1}_{i'_{j_5}i'_{j_6}}\right]_{15}\nonumber \\ 	
&&\hspace{0cm}+\left[\frac{1}{2^4}\sum_{j_1\dots j_6\in[1,\dots,6]}\tilde{\epsilon}_{j_1\dots j_6}\eta_{i'_{j_1}}\eta_{i'_{j_2}} S^{-1}_{i'_{j_3}i'_{j_4}} S^{-1}_{i'_{j_5}i'_{j_6}}\right]_{45} \nonumber \\
&&\hspace{0cm}-\left[\frac{1}{3! 2^3}\sum_{j_1\dots j_6\in[1,\dots,6]}\tilde{\epsilon}_{j_1\dots j_6} 	S^{-1}_{i'_{j_1}i'_{j_2}} S^{-1}_{i'_{j_3}i'_{j_4}} S^{-1}_{i'_{j_5}i'_{j_6}}\right]_{15}\nonumber   \,.
\end{eqnarray}
The expression for $K_{\rm B}$ inserting the three-point correlation function in the hierachical model   reads (see Eq.~\ref{eq:edge}): 
\begin{eqnarray}
\label{eq:kurt2} 
\mathcal K_{\rm B} (\mbi\nu)&\equiv& \frac{1}{6!}\sum_{i'j'k'l'm'n'}\nonumber\\
&&\hspace{-2.cm}\times\left[\frac{1}{3!3!2}\sum_{j_1\dots j_6\in[1,\dots,6]}\tilde{\epsilon}_{j_1\dots j_6}\langle\Phi_{i'_{j_1}}\Phi_{i'_{j_2}}\Phi_{i'_{j_3}}\rangle_{\rm c}\langle\Phi_{i'_{j_4}}\Phi_{i'_{j_5}}\Phi_{i'_{j_6}}\rangle_{\rm c}\right]_{10}\nonumber\\
&&\hspace{-2.cm}\times\tilde{h}_{i'j'k'l'm'n'}(\mbi \nu)\nonumber\\
&&\hspace{-1.5cm}=\frac{10}{6!}Q^2_3\sum_{i'j'k'l'm'n'}\left[ S_{i'j'} S_{i'k'}+ S_{i'j'} S_{j'k'}+ S_{i'k'} S_{j'k'}\right]\nonumber\\
&&\hspace{-2.cm}\times\left[ S_{l'm'} S_{l'n'}+ S_{l'm'} S_{m'n'}+ S_{l'n'} S_{m'n'}\right]\nonumber\\
&&\hspace{-2.cm}\times\tilde{h}_{i'j'k'l'm'n'}(\mbi \nu)\,.
\end{eqnarray}
After making the corresponding calculations (see appendix \ref{app:kurt2})  we find that the second contribution $\mathcal K_{\rm B}$  to the kurtosis term $\mathcal K$ can be written as\begin{eqnarray}
\label{eq:kurt22} 
\lefteqn{\mathcal K_{\rm B}(\mat S^{-1/2}\mbi \Phi)=}\\
&&\hspace{-0.5cm} Q_3^2\left[ \frac{1}{8}\left(\sum_{i}\Phi_{i}^2\eta_i\right)^2-\frac{1}{8}\sum_{ij}\Phi^2_{i}S_{ij}^{-1}\Phi^2_{j}-\frac{1}{2}\sum_{ij}\Phi_{i}\Phi_j^2\eta_{j}  \right.\nonumber\\
&&\hspace{-0.5cm}\left.  - \frac{1}{2} \sum_{i}\Phi^3_{i}\eta_i - \frac{1}{4} \sum_{ij}\Phi_i^2\eta_iS_{jj}\eta_j - \frac{1}{2}\sum_{ij}\eta_i\Phi_iS_{ij}\eta_j\Phi_j  \right.\nonumber\\
&&\hspace{-0.5cm}\left. + \frac{1}{4} \sum_{ij}S_{ii}S_{ij}^{-1}\Phi_j^2+\frac{1}{2}\sum_i\Phi_i^2+\frac{3}{2}\left(\sum_i\Phi_i \right)^2\right.\nonumber\\
&&\hspace{-0.5cm} \left.+2\sum_iS_{ii}\eta_i\Phi_i+\sum_{ij}S_{ij}\eta_j\Phi_j +\frac{1}{8} \left( \sum_iS_{ii}\eta_i\right)^2  \right.  \nonumber\\
&&\hspace{-0.5cm} \left. +\frac{1}{4} \sum_{ij}\eta_iS_{ij}^2\eta_j -\frac{3}{4} \sum_{ij}S_{ij}-\sum_{i}S_{ii}-\frac{1}{8}\sum_{ij}S_{ii}S_{ij}^{-1}S_{jj}  \right]  \nonumber\,.
\end{eqnarray}
We verify the skewness and kurtosis terms presented here by comparing the univariate case  Eqs.~(\ref{eq:skewUNI}, \ref{eq:kurt1UNI}, and \ref{eq:kurt2UNI})  to the  corresponding Eqs.~(\ref{eq:skewres}, \ref{eq:kurt1res}, and \ref{eq:kurt2res}) simplified to a single index  (see appendices \ref{app:UNI}, \ref{app:skew} and \ref{app:kurt}).
Now the first and second kurtosis contribution terms can be added and plugged in Eq.~(\ref{eq:edge}) together with the skewness found in the previous section to calculate the probability distribution function of the matter field.
We would like to emphasize here that the expressions found in this section for the skewness and the kurtosis can be efficiently computed (see Eqs.~\ref{eq:skew}, \ref{eq:kurt12} and \ref{eq:kurt22}). It should be noticed, that they rely on basic operations with the fft required to perform convolutions being the most expensive one.

\section{Summary and conclusions}

The avalanche of astronomical data coming from different surveys which scan different epochs of the Universe makes it possible to map the Universe with unprecedented accuracy.
 We stress here the necessity to model as precise as possible the matter statistics so that the multidimensional picture of the Universe can be reconstructed with the highest possible fidelity. This implies providing a multivariate higher order statistical description of the structures in the Universe.

We have extended the works done for the univariate case and presented the multivariate Edgeworth expansion of the lognormal field. We made the expansion explicitly up to third order in perturbation theory where we had to calculate the multivariate Hermite polynomials up to sixth order. The skewness and kurtosis terms include the two-point, the three-point, and the four-point correlation functions. 

We could show that these terms can be calculated using analytical expressions for the higher order correlation functions like the ones provided by the hierarchical model. 

The expressions derived in this work could be used to generate and reconstruct three-dimensional matter fields using higher order correlations  within a Bayesian framework applying for example the Hybrid Markov Chain Monte Carlo Hamiltonian sampling technique \cite[see the works by][]{jasche_hamil,kitaura_lyman}.
This could have interesting applications to study non-Gaussianity in the Large-Scale Structure. 

As new astronomical windows    are being opened to map the Universe at earlier times in which structures were not clustered as much as they are today, we think that the study of the moderate nonlinear regime is crucial for an
accurate data analysis.
Although numerical N-body simulations provide a magnificent tool to model structure formation, almost without resolution restrictions in comparison to the resolution provided by astronomical observations, they do not picture the actual realization of the Universe.
Therefore, we believe that a special effort should be done to model the multivariate statistical nature of the matter distribution which permits one to extract as much information as possible directly from the observational data.  We hope that this work serves to contribute in this direction.

\section*{Acknowledgments}

I thank Andrea Ferrara for encouraging scientific conversations and for providing me all the necessary support.   I  thank Rien Van de Weygaert for motivating discussions on the Cosmological Large-Scale Structure.  I thank especially Ignacio Cernuda for comments on the manuscript. I thank Euihun Joung and Sunghye Baek for useful discussions on integrals, quantum mechanics and numerical computing. 

Warm thanks to the organizers of the Astronomical Data Analysis 6th conference (3rd-6th May 2010) and the Cosmic Co-Motion Workshop (27th-30th Sep 2010) for letting me present part of this work.

The author thanks the {\it Intra-European Marie Curie fellowship} with project number 221783 and acronym {\it MCMCLYMAN} for supporting this project and {\it The Cluster of Excellence for Fundamental Physics on the Origin and Structure of the Universe} for supporting the final stage of this project.

{\small
\bibliographystyle{mn2e}
\bibliography{lit}

\begin{thebibliography}{}

\bibitem[\protect\citeauthoryear{{Abazajian}, {Adelman-McCarthy},
  {Ag{\"u}eros}, {Allam}, {Allende Prieto}, {An}, {Anderson} \&
  {Anderson}}{{Abazajian} et~al.}{2009}]{2009ApJS..182..543A}
{Abazajian} K.~N.,  {Adelman-McCarthy} J.~K.,  {Ag{\"u}eros} M.~A.,  {Allam}
  S.~S.,  {Allende Prieto} C.,  {An} D.,  {Anderson} K.~S.~J.,    {Anderson}
  S.~F.,  2009, \apjs, 182, 543

\bibitem[\protect\citeauthoryear{{Albrecht} \& {Steinhardt}}{{Albrecht} \&
  {Steinhardt}}{1982}]{1982PhRvL..48.1220A}
{Albrecht} A.,  {Steinhardt} P.~J.,  1982, Physical Review Letters, 48, 1220

\bibitem[\protect\citeauthoryear{{Arag{\'o}n-Calvo}, {Jones}, {van de Weygaert}
  \& {van der Hulst}}{{Arag{\'o}n-Calvo} et~al.}{2007}]{2007A&A...474..315A}
{Arag{\'o}n-Calvo} M.~A.,  {Jones} B.~J.~T.,  {van de Weygaert} R.,    {van der
  Hulst} J.~M.,  2007, \aap, 474, 315

\bibitem[\protect\citeauthoryear{{Balian} \& {Schaeffer}}{{Balian} \&
  {Schaeffer}}{1989}]{1989A&A...220....1B}
{Balian} R.,  {Schaeffer} R.,  1989, \aap, 220, 1

\bibitem[\protect\citeauthoryear{{Bardeen}, {Bond}, {Kaiser} \&
  {Szalay}}{{Bardeen} et~al.}{1986}]{1986ApJ...304...15B}
{Bardeen} J.~M.,  {Bond} J.~R.,  {Kaiser} N.,    {Szalay} A.~S.,  1986, \apj,
  304, 15

\bibitem[\protect\citeauthoryear{{Bardeen}, {Steinhardt} \& {Turner}}{{Bardeen}
  et~al.}{1983}]{1983PhRvD..28..679B}
{Bardeen} J.~M.,  {Steinhardt} P.~J.,    {Turner} M.~S.,  1983, \prd, 28, 679

\bibitem[\protect\citeauthoryear{{Barndorff-Nielsen} \&
  {Cox}}{{Barndorff-Nielsen} \& {Cox}}{1989}]{barndorff}
{Barndorff-Nielsen} H.,  {Cox} D.~R.,  1989, {Asymptotic techniques for use in
  statistics}.
Chapman \& Hall, London

\bibitem[\protect\citeauthoryear{Berkowitz \& Garner}{Berkowitz \&
  Garner}{1970}]{Berkowitz:1970:CMH}
Berkowitz S.,  Garner F.~J.,  1970, Mathematics of Computation, 24, 537

\bibitem[\protect\citeauthoryear{{Bernardeau}, {Colombi}, {Gazta{\~n}aga} \&
  {Scoccimarro}}{{Bernardeau} et~al.}{2002}]{2002PhR...367....1B}
{Bernardeau} F.,  {Colombi} S.,  {Gazta{\~n}aga} E.,    {Scoccimarro} R.,
  2002, \physrep, 367, 1

\bibitem[\protect\citeauthoryear{{Bernardeau} \& {Kofman}}{{Bernardeau} \&
  {Kofman}}{1995}]{1995ApJ...443..479B}
{Bernardeau} F.,  {Kofman} L.,  1995, \apj, 443, 479

\bibitem[\protect\citeauthoryear{{Blinnikov} \& {Moessner}}{{Blinnikov} \&
  {Moessner}}{1998}]{1998A&AS..130..193B}
{Blinnikov} S.,  {Moessner} R.,  1998, \aaps, 130, 193

\bibitem[\protect\citeauthoryear{{Bouchet}, {Colombi}, {Hivon} \&
  {Juszkiewicz}}{{Bouchet} et~al.}{1995}]{bouchet1995}
{Bouchet} F.~R.,  {Colombi} S.,  {Hivon} E.,    {Juszkiewicz} R.,  1995, \aap,
  296, 575

\bibitem[\protect\citeauthoryear{{Buchert}}{{Buchert}}{1994}]{1994MNRAS.267..8%
11B}
{Buchert} T.,  1994, \mnras, 267, 811

\bibitem[\protect\citeauthoryear{{Buchert} \& {Ehlers}}{{Buchert} \&
  {Ehlers}}{1993}]{1993MNRAS.264..375B}
{Buchert} T.,  {Ehlers} J.,  1993, \mnras, 264, 375

\bibitem[\protect\citeauthoryear{{Bunn}, {Fisher}, {Hoffman}, {Lahav}, {Silk}
  \& {Zaroubi}}{{Bunn} et~al.}{1994}]{1994ApJ...432L..75B}
{Bunn} E.~F.,  {Fisher} K.~B.,  {Hoffman} Y.,  {Lahav} O.,  {Silk} J.,
  {Zaroubi} S.,  1994, \apjl, 432, L75

\bibitem[\protect\citeauthoryear{{Chen}, {Vergely}, {Valette} \&
  {Carraro}}{{Chen} et~al.}{1998}]{1998A&A...336..137C}
{Chen} B.,  {Vergely} J.~L.,  {Valette} B.,    {Carraro} G.,  1998, \aap, 336,
  137

\bibitem[\protect\citeauthoryear{{Coles} \& {Jones}}{{Coles} \&
  {Jones}}{1991}]{1991MNRAS.248....1C}
{Coles} P.,  {Jones} B.,  1991, \mnras, 248, 1

\bibitem[\protect\citeauthoryear{{Colless}, {Peterson}, {Jackson}, {Peacock},
  {Cole}, {Norberg}, {Baldry} \& {Baugh}}{{Colless}
  et~al.}{2003}]{2003astro.ph..6581C}
{Colless} M.,  {Peterson} B.~A.,  {Jackson} C.,  {Peacock} J.~A.,  {Cole} S.,
  {Norberg} P.,  {Baldry} I.~K.,    {Baugh} C.~M.,  2003, ArXiv Astrophysics
  e-prints

\bibitem[\protect\citeauthoryear{{Colombi}}{{Colombi}}{1994}]{colombi}
{Colombi} S.,  1994, \apj, 435, 536

\bibitem[\protect\citeauthoryear{{Contaldi}, {Ferreira}, {Magueijo} \&
  {G{\'o}rski}}{{Contaldi} et~al.}{2000}]{2000ApJ...534...25C}
{Contaldi} C.~R.,  {Ferreira} P.~G.,  {Magueijo} J.,    {G{\'o}rski} K.~M.,
  2000, \apj, 534, 25

\bibitem[\protect\citeauthoryear{{Contaldi} \& {Magueijo}}{{Contaldi} \&
  {Magueijo}}{2001}]{2001PhRvD..63j3512C}
{Contaldi} C.~R.,  {Magueijo} J.,  2001, \prd, 63, 103512

\bibitem[\protect\citeauthoryear{{Cramer}}{{Cramer}}{1946}]{cramer}
{Cramer} H.,  1946, {Mathematical methods of statistics}.
Princeton University Press, 1946.~575 p.

\bibitem[\protect\citeauthoryear{{Crotts}, {Garnavich}, {Priedhorsky}, {Habib},
  {Heitmann}, {Wang}, {Baron} \& {Branch}}{{Crotts}
  et~al.}{2005}]{2005astro.ph..7043C}
{Crotts} A.,  {Garnavich} P.,  {Priedhorsky} W.,  {Habib} S.,  {Heitmann} K.,
  {Wang} Y.,  {Baron} E.,    {Branch} D.,  2005, ArXiv Astrophysics e-prints

\bibitem[\protect\citeauthoryear{{Davis}, {Faber}, {Newman}, {Phillips},
  {Ellis}, {Steidel}, {Conselice} \& {Coil}}{{Davis}
  et~al.}{2003}]{2003SPIE.4834..161D}
{Davis} M.,  {Faber} S.~M.,  {Newman} J.,  {Phillips} A.~C.,  {Ellis} R.~S.,
  {Steidel} C.~C.,  {Conselice} C.,    {Coil} A.~L.,  2003, in
  {P.~Guhathakurta} ed., Society of Photo-Optical Instrumentation Engineers
  (SPIE) Conference Series Vol.~4834 of Society of Photo-Optical
  Instrumentation Engineers (SPIE) Conference Series, {Science Objectives and
  Early Results of the DEEP2 Redshift Survey}.
pp 161--172

\bibitem[\protect\citeauthoryear{{Erdo{\u g}du}, {Lahav}, {Huchra} \& {et
  al.}}{{Erdo{\u g}du} et~al.}{2006}]{2006MNRAS.373...45E}
{Erdo{\u g}du} P.,  {Lahav} O.,  {Huchra} J.,    {et al.} 2006, \mnras, 373, 45

\bibitem[\protect\citeauthoryear{{Erdo{\u g}du}, {Lahav}, {Zaroubi},
  {Efstathiou}, {Moody}, {Peacock}, {Colless}, {Baldry} \& {et al.}}{{Erdo{\u
  g}du} et~al.}{2004}]{2004MNRAS.352..939E}
{Erdo{\u g}du} P.,  {Lahav} O.,  {Zaroubi} S.,  {Efstathiou} G.,  {Moody} S.,
  {Peacock} J.~A.,  {Colless} M.,  {Baldry} I.~K.,    {et al.} 2004, \mnras,
  352, 939

\bibitem[\protect\citeauthoryear{{Eriksen}, {Huey}, {Saha}, {Hansen}, {Dick},
  {Banday}, {G{\'o}rski}, {Jain}, {Jewell}, {Knox}, {Larson}, {O'Dwyer},
  {Souradeep} \& {Wandelt}}{{Eriksen} et~al.}{2007}]{2007ApJ...656..641E}
{Eriksen} H.~K.,  {Huey} G.,  {Saha} R.,  {Hansen} F.~K.,  {Dick} J.,  {Banday}
  A.~J.,  {G{\'o}rski} K.~M.,  {Jain} P.,  {Jewell} J.~B.,  {Knox} L.,
  {Larson} D.~L.,  {O'Dwyer} I.~J.,  {Souradeep} T.,    {Wandelt} B.~D.,  2007,
  \apj, 656, 641

\bibitem[\protect\citeauthoryear{{Falcke}, {van Haarlem}, {de Bruyn}, {Braun},
  {R{\"o}ttgering}, {Stappers}, {Boland}, {Butcher} \& {et al}}{{Falcke}
  et~al.}{2007}]{2007HiA....14..386F}
{Falcke} H.~D.,  {van Haarlem} M.~P.,  {de Bruyn} A.~G.,  {Braun} R.,
  {R{\"o}ttgering} H.~J.~A.,  {Stappers} B.,  {Boland} W.~H.~W.~M.,  {Butcher}
  H.~R.,    {et al} 2007, Highlights of Astronomy, 14, 386

\bibitem[\protect\citeauthoryear{{Fisher}, {Lahav}, {Hoffman}, {Lynden-Bell} \&
  {Zaroubi}}{{Fisher} et~al.}{1995}]{1995MNRAS.272..885F}
{Fisher} K.~B.,  {Lahav} O.,  {Hoffman} Y.,  {Lynden-Bell} D.,    {Zaroubi} S.,
   1995, \mnras, 272, 885

\bibitem[\protect\citeauthoryear{{Fry}}{{Fry}}{1984}]{1984ApJ...277L...5F}
{Fry} J.~N.,  1984, \apjl, 277, L5

\bibitem[\protect\citeauthoryear{{Fry}}{{Fry}}{1986}]{1986ApJ...306..358F}
{Fry} J.~N.,  1986, \apj, 306, 358

\bibitem[\protect\citeauthoryear{{Fry} \& {Peebles}}{{Fry} \&
  {Peebles}}{1978}]{1978ApJ...221...19F}
{Fry} J.~N.,  {Peebles} P.~J.~E.,  1978, \apj, 221, 19

\bibitem[\protect\citeauthoryear{{Gazta{\~n}aga}, {Fosalba} \&
  {Elizalde}}{{Gazta{\~n}aga} et~al.}{2000}]{2000ApJ...539..522G}
{Gazta{\~n}aga} E.,  {Fosalba} P.,    {Elizalde} E.,  2000, \apj, 539, 522

\bibitem[\protect\citeauthoryear{{Gott} III, {Hambrick}, {Vogeley}, {Kim},
  {Park}, {Choi}, {Cen}, {Ostriker} \& {Nagamine}}{{Gott}
  et~al.}{2008}]{2008ApJ...675...16G}
{Gott} III J.~R.,  {Hambrick} D.~C.,  {Vogeley} M.~S.,  {Kim} J.,  {Park} C.,
  {Choi} Y.,  {Cen} R.,  {Ostriker} J.~P.,    {Nagamine} K.,  2008, \apj, 675,
  16

\bibitem[\protect\citeauthoryear{{Guth}}{{Guth}}{1981}]{1981PhRvD..23..347G}
{Guth} A.~H.,  1981, \prd, 23, 347

\bibitem[\protect\citeauthoryear{{Guth} \& {Pi}}{{Guth} \&
  {Pi}}{1982}]{1982PhRvL..49.1110G}
{Guth} A.~H.,  {Pi} S.-Y.,  1982, Physical Review Letters, 49, 1110

\bibitem[\protect\citeauthoryear{{Hawking}}{{Hawking}}{1982}]{1982CMaPh..87..3%
95H}
{Hawking} S.~W.,  1982, Communications in Mathematical Physics, 87, 395

\bibitem[\protect\citeauthoryear{{Hoekstra}, {Mellier}, {van Waerbeke},
  {Semboloni}, {Fu}, {Hudson}, {Parker}, {Tereno} \& {Benabed}}{{Hoekstra}
  et~al.}{2006}]{2006ApJ...647..116H}
{Hoekstra} H.,  {Mellier} Y.,  {van Waerbeke} L.,  {Semboloni} E.,  {Fu} L.,
  {Hudson} M.~J.,  {Parker} L.~C.,  {Tereno} I.,    {Benabed} K.,  2006, \apj,
  647, 116

\bibitem[\protect\citeauthoryear{{Hubble}}{{Hubble}}{1934}]{1934ApJ....79....8%
H}
{Hubble} E.,  1934, \apj, 79, 8

\bibitem[\protect\citeauthoryear{{James}, {Colless}, {Lewis} \&
  {Peacock}}{{James} et~al.}{2009}]{2009MNRAS.394..454J}
{James} J.~B.,  {Colless} M.,  {Lewis} G.~F.,    {Peacock} J.~A.,  2009,
  \mnras, 394, 454

\bibitem[\protect\citeauthoryear{{Jasche} \& {Kitaura}}{{Jasche} \&
  {Kitaura}}{2010}]{jasche_hamil}
{Jasche} J.,  {Kitaura} F.~S.,  2010, \mnras, 407, 29

\bibitem[\protect\citeauthoryear{{Jasche}, {Kitaura}, {Li} \&
  {En{\ss}lin}}{{Jasche} et~al.}{2010}]{jasche_sdss}
{Jasche} J.,  {Kitaura} F.~S.,  {Li} C.,    {En{\ss}lin} T.~A.,  2010, \mnras,
  pp 1638--+

\bibitem[\protect\citeauthoryear{{Juszkiewicz}, {Weinberg}, {Amsterdamski},
  {Chodorowski} \& {Bouchet}}{{Juszkiewicz} et~al.}{1995}]{1995ApJ...442...39J}
{Juszkiewicz} R.,  {Weinberg} D.~H.,  {Amsterdamski} P.,  {Chodorowski} M.,
  {Bouchet} F.,  1995, \apj, 442, 39

\bibitem[\protect\citeauthoryear{{Kaiser} \& {Pan-STARRS Team}}{{Kaiser} \&
  {Pan-STARRS Team}}{2002}]{2002AAS...20112207K}
{Kaiser} N.,  {Pan-STARRS Team} 2002, in Bulletin of the American Astronomical
  Society Vol.~34 of Bulletin of the American Astronomical Society, {The
  Pan-STARRS Optical Survey Telescope Project}.
pp 1304--+

\bibitem[\protect\citeauthoryear{{Kitaura}, {Gallerani} \& {Ferrara}}{{Kitaura}
  et~al.}{2010}]{kitaura_lyman}
{Kitaura} F.~S.,  {Gallerani} S.,    {Ferrara} A.,  2010, ArXiv e-prints

\bibitem[\protect\citeauthoryear{{Kitaura}, {Jasche}, {Li}, {En{\ss}lin},
  {Metcalf}, {Wandelt}, {Lemson} \& {White}}{{Kitaura}
  et~al.}{2009}]{kitaura_sdss}
{Kitaura} F.~S.,  {Jasche} J.,  {Li} C.,  {En{\ss}lin} T.~A.,  {Metcalf} R.~B.,
   {Wandelt} B.~D.,  {Lemson} G.,    {White} S.~D.~M.,  2009, \mnras, 400, 183

\bibitem[\protect\citeauthoryear{{Kitaura}, {Jasche} \& {Metcalf}}{{Kitaura}
  et~al.}{2010}]{kitaura_log}
{Kitaura} F.~S.,  {Jasche} J.,    {Metcalf} R.~B.,  2010, \mnras, 403, 589

\bibitem[\protect\citeauthoryear{{Le F{\`e}vre}, {Vettolani}, {Paltani},
  {Tresse}, {Zamorani}, {Le Brun}, {Moreau} \& {Bottini}}{{Le F{\`e}vre}
  et~al.}{2004}]{2004A&A...428.1043L}
{Le F{\`e}vre} O.,  {Vettolani} G.,  {Paltani} S.,  {Tresse} L.,  {Zamorani}
  G.,  {Le Brun} V.,  {Moreau} C.,    {Bottini} D.,  2004, \aap, 428, 1043

\bibitem[\protect\citeauthoryear{{Linde}}{{Linde}}{1982}]{1982PhLB..108..389L}
{Linde} A.~D.,  1982, Physics Letters B, 108, 389

\bibitem[\protect\citeauthoryear{{Lonsdale}, {Cappallo}, {Morales}, {Briggs},
  {Benkevitch}, {Bowman}, {Bunton}, {Burns} \& {et al}}{{Lonsdale}
  et~al.}{2009}]{2009IEEEP..97.1497L}
{Lonsdale} C.~J.,  {Cappallo} R.~J.,  {Morales} M.~F.,  {Briggs} F.~H.,
  {Benkevitch} L.,  {Bowman} J.~D.,  {Bunton} J.~D.,  {Burns} S.,    {et al}
  2009, IEEE Proceedings, 97, 1497

\bibitem[\protect\citeauthoryear{{Matsubara}}{{Matsubara}}{2003}]{2003ApJ...58%
4....1M}
{Matsubara} T.,  2003, \apj, 584, 1

\bibitem[\protect\citeauthoryear{{Matsubara}}{{Matsubara}}{2008}]{2008PhRvD..7%
8h3519M}
{Matsubara} T.,  2008, \prd, 78, 083519

\bibitem[\protect\citeauthoryear{{Matsubara} \& {Suto}}{{Matsubara} \&
  {Suto}}{1994}]{1994ApJ...420..497M}
{Matsubara} T.,  {Suto} Y.,  1994, \apj, 420, 497

\bibitem[\protect\citeauthoryear{{Matsubara} \& {Yokoyama}}{{Matsubara} \&
  {Yokoyama}}{1996}]{1996ApJ...463..409M}
{Matsubara} T.,  {Yokoyama} J.,  1996, \apj, 463, 409

\bibitem[\protect\citeauthoryear{{McDonald}, {Seljak}, {Cen}, {Shih},
  {Weinberg}, {Burles}, {Schneider}, {Schlegel}, {Bahcall}, {Briggs},
  {Brinkmann}, {Fukugita}, {Ivezi{\'c}}, {Kent} \& {Vanden Berk}}{{McDonald}
  et~al.}{2005}]{2005ApJ...635..761M}
{McDonald} P.,  {Seljak} U.,  {Cen} R.,  {Shih} D.,  {Weinberg} D.~H.,
  {Burles} S.,  {Schneider} D.~P.,  {Schlegel} D.~J.,  {Bahcall} N.~A.,
  {Briggs} J.~W.,  {Brinkmann} J.,  {Fukugita} M.,  {Ivezi{\'c}} {\v Z}.,
  {Kent} S.,    {Vanden Berk} D.~E.,  2005, \apj, 635, 761

\bibitem[\protect\citeauthoryear{{Miralda-Escud{\'e}}, {Haehnelt} \&
  {Rees}}{{Miralda-Escud{\'e}} et~al.}{2000}]{2000ApJ...530....1M}
{Miralda-Escud{\'e}} J.,  {Haehnelt} M.,    {Rees} M.~J.,  2000, \apj, 530, 1

\bibitem[\protect\citeauthoryear{{Padmanabhan} \& {Subramanian}}{{Padmanabhan}
  \& {Subramanian}}{1993}]{1993ApJ...410..482P}
{Padmanabhan} T.,  {Subramanian} K.,  1993, \apj, 410, 482

\bibitem[\protect\citeauthoryear{{Parsons}, {Backer}, {Foster}, {Wright},
  {Bradley}, {Gugliucci}, {Parashare}, {Benoit} \& {et al}}{{Parsons}
  et~al.}{2010}]{2010AJ....139.1468P}
{Parsons} A.~R.,  {Backer} D.~C.,  {Foster} G.~S.,  {Wright} M.~C.~H.,
  {Bradley} R.~F.,  {Gugliucci} N.~E.,  {Parashare} C.~R.,  {Benoit} E.~E.,
  {et al} 2010, \aj, 139, 1468

\bibitem[\protect\citeauthoryear{{Pen}, {Chang}, {Peterson}, {Roy}, {Gupta} \&
  {Bandura}}{{Pen} et~al.}{2008}]{2008AIPC.1035...75P}
{Pen} U.,  {Chang} T.,  {Peterson} J.~B.,  {Roy} J.,  {Gupta} Y.,    {Bandura}
  K.,  2008, in {R.~Minchin \& E.~Momjian} ed., The Evolution of Galaxies
  Through the Neutral Hydrogen Window Vol.~1035 of American Institute of
  Physics Conference Series, {The GMRT Search for Reionization}.
pp 75--81

\bibitem[\protect\citeauthoryear{{Pichon}, {Vergely}, {Rollinde}, {Colombi} \&
  {Petitjean}}{{Pichon} et~al.}{2001}]{pichon}
{Pichon} C.,  {Vergely} J.~L.,  {Rollinde} E.,  {Colombi} S.,    {Petitjean}
  P.,  2001, \mnras, 326, 597

\bibitem[\protect\citeauthoryear{{R{\'e}fr{\'e}gier}, {Boulade}, {Mellier},
  {Milliard}, {Pain}, {Michaud}, {Safa} \& {Amara}}{{R{\'e}fr{\'e}gier}
  et~al.}{2006}]{2006SPIE.6265E..58R}
{R{\'e}fr{\'e}gier} A.,  {Boulade} O.,  {Mellier} Y.,  {Milliard} B.,  {Pain}
  R.,  {Michaud} J.,  {Safa} F.,    {Amara} A.,  2006, in Society of
  Photo-Optical Instrumentation Engineers (SPIE) Conference Series Vol.~6265 of
  Society of Photo-Optical Instrumentation Engineers (SPIE) Conference Series,
  {DUNE: the Dark Universe Explorer}

\bibitem[\protect\citeauthoryear{{Schaap} \& {van de Weygaert}}{{Schaap} \&
  {van de Weygaert}}{2000}]{2000A&A...363L..29S}
{Schaap} W.~E.,  {van de Weygaert} R.,  2000, \aap, 363, L29

\bibitem[\protect\citeauthoryear{{Scherrer} \& {Bertschinger}}{{Scherrer} \&
  {Bertschinger}}{1991}]{1991ApJ...381..349S}
{Scherrer} R.~J.,  {Bertschinger} E.,  1991, \apj, 381, 349

\bibitem[\protect\citeauthoryear{{Schlegel}, {White} \&
  {Eisenstein}}{{Schlegel} et~al.}{2009}]{2009astro2010S.314S}
{Schlegel} D.,  {White} M.,    {Eisenstein} D.,  2009, in astro2010: The
  Astronomy and Astrophysics Decadal Survey Vol.~2010 of Astronomy, {The Baryon
  Oscillation Spectroscopic Survey: Precision measurement of the absolute
  cosmic distance scale}.
pp 314--+

\bibitem[\protect\citeauthoryear{{Scoccimarro}, {Colombi}, {Fry}, {Frieman},
  {Hivon} \& {Melott}}{{Scoccimarro} et~al.}{1998}]{1998ApJ...496..586S}
{Scoccimarro} R.,  {Colombi} S.,  {Fry} J.~N.,  {Frieman} J.~A.,  {Hivon} E.,
   {Melott} A.,  1998, \apj, 496, 586

\bibitem[\protect\citeauthoryear{{Starobinsky}}{{Starobinsky}}{1982}]{1982PhLB%
..117..175S}
{Starobinsky} A.~A.,  1982, Physics Letters B, 117, 175

\bibitem[\protect\citeauthoryear{{Suto} \& {Matsubara}}{{Suto} \&
  {Matsubara}}{1994}]{1994ApJ...420..504S}
{Suto} Y.,  {Matsubara} T.,  1994, \apj, 420, 504

\bibitem[\protect\citeauthoryear{{Szapudi}, {Colombi}, {Jenkins} \&
  {Colberg}}{{Szapudi} et~al.}{2000}]{2000MNRAS.313..725S}
{Szapudi} I.,  {Colombi} S.,  {Jenkins} A.,    {Colberg} J.,  2000, \mnras,
  313, 725

\bibitem[\protect\citeauthoryear{{Tarantola} \& {Valette}}{{Tarantola} \&
  {Valette}}{1982}]{tarantola}
{Tarantola} A.,  {Valette} B.,  1982, Reviews of Geophysics and Space Physics,
  20, 219

\bibitem[\protect\citeauthoryear{{Taruya} \& {Soda}}{{Taruya} \&
  {Soda}}{1999}]{1999ApJ...522...46T}
{Taruya} A.,  {Soda} J.,  1999, \apj, 522, 46

\bibitem[\protect\citeauthoryear{{Tegmark}}{{Tegmark}}{1997}]{1997ApJ...480L..%
87T}
{Tegmark} M.,  1997, \apjl, 480, L87+

\bibitem[\protect\citeauthoryear{{The Dark Energy Survey Collaboration}}{{The
  Dark Energy Survey Collaboration}}{2005}]{2005astro.ph.10346T}
{The Dark Energy Survey Collaboration} 2005, ArXiv Astrophysics e-prints

\bibitem[\protect\citeauthoryear{{van de Weygaert}, {Aragon-Calvo}, {Jones} \&
  {Platen}}{{van de Weygaert} et~al.}{2009}]{2009arXiv0912.3448V}
{van de Weygaert} R.,  {Aragon-Calvo} M.~A.,  {Jones} B.~J.~T.,    {Platen} E.,
   2009, ArXiv e-prints

\bibitem[\protect\citeauthoryear{{Webster}, {Lahav} \& {Fisher}}{{Webster}
  et~al.}{1997}]{1997MNRAS.287..425W}
{Webster} M.,  {Lahav} O.,    {Fisher} K.,  1997, \mnras, 287, 425

\bibitem[\protect\citeauthoryear{{Wild}, {Peacock}, {Lahav}, {Conway},
  {Maddox}, {Baldry}, {Baugh} \& et al}{{Wild}
  et~al.}{2005}]{2005MNRAS.356..247W}
{Wild} V.,  {Peacock} J.~A.,  {Lahav} O.,  {Conway} E.,  {Maddox} S.,  {Baldry}
  I.~K.,  {Baugh} C.~M.,    et al 2005, \mnras, 356, 247

\bibitem[\protect\citeauthoryear{{Zaroubi}, {Hoffman} \& {Dekel}}{{Zaroubi}
  et~al.}{1999}]{1999ApJ...520..413Z}
{Zaroubi} S.,  {Hoffman} Y.,    {Dekel} A.,  1999, \apj, 520, 413

\bibitem[\protect\citeauthoryear{{Zaroubi}, {Hoffman}, {Fisher} \&
  {Lahav}}{{Zaroubi} et~al.}{1995}]{zaroubi}
{Zaroubi} S.,  {Hoffman} Y.,  {Fisher} K.~B.,    {Lahav} O.,  1995, \apj, 449,
  446

\bibitem[\protect\citeauthoryear{{Zel'dovich}}{{Zel'dovich}}{1970}]{1970A&A...%
..5...84Z}
{Zel'dovich} Y.~B.,  1970, \aap, 5, 84

\bibitem[\protect\citeauthoryear{{Zheng}}{{Zheng}}{2004}]{2004ApJ...614..527Z}
{Zheng} Z.,  2004, \apj, 614, 527

\end{thebibliography}
}

\clearpage

\pagebreak
        
\appendix

\section{Univariate skewness and kurtosis terms in the hierarchical model}
\label{app:UNI}

From the hierarchical model ansatz (see section \ref{sec:HM}) we can define the univariate three-point correlation function as: $\xi_3=3 Q_3 \sigma^4$ and the four-point correlation function by: $\xi_4=4 Q_4^{\rm a} \sigma^6+12 Q_4^{\rm b} \sigma^6$.
Let us now look at the skewness and kurtosis terms.

\subsection{Skewness in the univariate case}
\label{app:skewUNI}

For the skewness we need to define the third order univariate weighted Hermite polynomial: $\tilde{h}_3(\sigma^{-1}\Phi)\equiv \sigma^{-3}h_3(\sigma^{-1}\Phi)=\sigma^{-6}\Phi^3-3\sigma^{-4}\Phi$.
We then get:
\begin{equation}
\label{eq:skewUNI}
\mathcal S(\sigma^{-1}\Phi)=\frac{1}{3!}\kappa_3h_3(\sigma^{-1}\Phi)=\frac{1}{3!}\xi_3\tilde{h}_3(\sigma^{-1}\Phi)=\frac{Q_3}{2}(\sigma^{-2}\Phi^3-3 \Phi)\,.
\end{equation}

\subsection{Kurtosis in the univariate case}
\label{app:kurtUNI}

In the case of the kurtosis terms we need to define the fourth and sixth order univariate weighted Hermite polynomials: $\tilde{h}_4(\sigma^{-1}\Phi)\equiv \sigma^{-4}h_4(\sigma^{-1}\Phi)=\sigma^{-8}\Phi^4-6\sigma^{-6}\Phi^2+3\sigma^{-4}$ and $\tilde{h}_6(\sigma^{-1}\Phi)\equiv \sigma^{-6}h_6(\sigma^{-1}\Phi)=\sigma^{-12}\Phi^6-15\sigma^{-10}\Phi^4+45\sigma^{-8}\Phi^2-15\sigma^{-6}$.

\subsubsection{First group of kurtosis terms}

Looking at the fourth order Hermite polynomial contribution we get:
\begin{eqnarray}
\label{eq:kurt1UNI}
\lefteqn{\mathcal K_{\rm A}(\sigma^{-1}\Phi)=\frac{1}{4!}\kappa_4h_4(\sigma^{-1}\Phi)=\frac{1}{4!}\xi_4\tilde{h}_4(\sigma^{-1}\Phi)}\\
&&=\frac{1}{6}\left(Q^{\rm a}_4+3Q_4^{\rm b}\right)(\sigma^{-2}\Phi^4-6 \Phi^2+3\sigma^2)\nonumber\,.
\end{eqnarray}

\subsubsection{Second group of kurtosis terms}

We then get:
\begin{eqnarray}
\label{eq:kurt2UNI}
\lefteqn{\mathcal K_{\rm B}(\sigma^{-1}\Phi)=\frac{10}{6!}\kappa_3^2h_6(\sigma^{-1}\Phi)=\frac{10}{6!}\xi_3^2\tilde{h}_6(\sigma^{-1}\Phi)}\\
&&=\frac{Q^2_3}{8}(\sigma^{-4}\Phi^6-15\sigma^{-2}\Phi^4+45\Phi^2-15\sigma^2)\nonumber\,.
\end{eqnarray}

\section{Univariate mean and variance}
\label{app:univar}

Here we will derive the univariate relations for the mean and variance of the expanded lognormal distribution. Note, that the results presented here are in agreement with the general formula presented by \citet[][]{colombi} without an explicit derivation.
Let us define the $k$-th order characteristic function for the variable $\Phi$ as a function of $t$:
\begin{equation}
\mathcal M_{\Phi k}(t)\equiv\langle e^{t\Phi}\rangle_k=\int\dd \Phi\,P_k(\Phi)e^{t\Phi}\,.
\end{equation}
We will look at this expression for the different orders and derive from it the corresponding mean and variance.

\subsection{Case $k=1$: lognormal}
\label{app:k1}

One finds by shifting the Gaussian integral the solution to the characteristic function of the lognormal distribution to be given by:
\begin{eqnarray}
\label{eq:m1}
\lefteqn{\mathcal M_{\Phi 1}(t)\equiv\langle e^{t\Phi}\rangle_1=\frac{1}{\sqrt{2\pi}\sigma}\int\dd \Phi\,e^{-\frac{1}{2}\frac{\Phi^2}{\sigma^2}+t\Phi}}\nonumber\\
&&=e^{\frac{t^2}{2}\sigma^2}\frac{1}{\sqrt{2\pi}\sigma}\int\dd \Phi\,e^{-\frac{1}{2}\frac{(\Phi-\sigma^2t)^2}{\sigma^2}}=e^{\frac{t^2}{2}\sigma^2}\,,
\end{eqnarray}
where we use the following definition: $\sigma^2\equiv\langle\Phi^2\rangle_k$.
Recalling the expression for $\Phi$: $\Phi\equiv\log(\frac{\rho}{\overline{\rho}})-\mu_s$ with $\overline{\rho}\equiv\langle\rho\rangle_k$ and $\mu_s=\langle s\rangle_k$, we can write the density as: $\rho=\overline{\rho}e^{\Phi+\mu_s}$.
Using the characteristic function we can calculate all the  moments of the density:
\begin{equation}
\langle\rho^t\rangle_k=\int\dd \Phi\,P_k(\Phi)\rho^t=\overline{\rho}^te^{t\mu_s}\langle e^{t\Phi}\rangle_k\,.
\end{equation}
For the lognormal distribution we have: 
\begin{equation}
\langle\rho^t\rangle_1=\frac{1}{\sqrt{2\pi}\sigma}\int\dd \Phi\,e^{-\frac{1}{2}\frac{\Phi^2}{\sigma^2}}\rho^t=\overline{\rho}^te^{t\mu_s}\langle e^{t\Phi}\rangle_1=\overline{\rho}^te^{t\mu_s+\frac{t^2}{2}\sigma^2}\,.
\end{equation}
\subsubsection{Mean}
The particular case for $t=1$: 
\begin{equation}
\langle\rho\rangle_1=\overline{\rho}e^{\mu_s+\frac{1}{2}\sigma^2}\,,
\end{equation}
leads to the mean:
\begin{equation}
\mu_s=-\frac{1}{2}\sigma^2\,.
\end{equation}
\subsubsection{Variance}
Whereas the case $t=2$:
\begin{equation}
\langle\rho^2\rangle_1=\overline{\rho}^2e^{2\mu_s+2\sigma^2}\,,
\end{equation}
relates the variance of the overdensity  $\delta_{\rm M}$ to the variance of $\Phi$:
\begin{equation}
\frac{\langle\rho^2\rangle_1}{\overline{\rho}^2}=\langle\delta_{\rm M}^2\rangle_1+1=e^{\sigma^2}\,.
\end{equation}
Thus, one finds that the variance $\sigma^2$ of $\Phi$ under the lognormal assumption is given by: 
\begin{equation}
\sigma^2=\ln(\sigma_\delta^2+1)\,,
\end{equation}
with $\sigma_\delta^2\equiv\langle\delta_{\rm M}^2\rangle_k$.

\subsection{Case $k=2$: lognormal with skewness}
\label{app:k2}

The characteristic function including skewness yields:
\begin{eqnarray}
\lefteqn{\mathcal M_{\Phi 2}(t)\equiv\langle e^{t\Phi}\rangle_2}\\
&&\hspace{-0cm}=\frac{1}{\sqrt{2\pi}\sigma}\int\dd \Phi\,e^{-\frac{1}{2}\frac{\Phi^2}{\sigma^2}}\left(1+\frac{1}{3!}\xi_3\tilde{h}_3\right)e^{t\Phi}\nonumber \\
&&\hspace{-0cm}=\frac{1}{\sqrt{2\pi}\sigma}\int\dd \Phi\,e^{-\frac{1}{2}\frac{\Phi^2}{\sigma^2}}\left(1+\frac{1}{3!}\xi_3\sigma^{-4}(\sigma^{-2}\Phi^3-3\Phi)\right)e^{t\Phi}\nonumber\,.
\end{eqnarray}
We have to solve integrals including an exponential term and polynomials of $\Phi$. Note however, that we can generate such integrals by subsequent derivatives of the characteristic function:
\begin{equation}
\langle e^{t\Phi}\Phi^n\rangle_1=\frac{\dd^n}{\dd t^n}\mathcal M_{\Phi 1}(t)=\frac{1}{\sqrt{2\pi}\sigma}\int\dd \Phi\,e^{-\frac{1}{2}\frac{\Phi^2}{\sigma^2}+t\Phi}\Phi^n\,.
\end{equation}
Taking the expression of the generating function for $k=1$ (Eq:~\ref{eq:m1}) we can calculate the solutions to the integrals including different powers of $\Phi$:
\begin{eqnarray}
\frac{\dd}{\dd t}\mathcal M_{\Phi 1}(t)&=&t\sigma^2e^{\frac{t^2}{2}\sigma^2}\nonumber   \\
\frac{\dd^2}{\dd t^2}\mathcal M_{\Phi 1}(t) &=&\left(t^2\sigma^2+1\right)\sigma^2e^{\frac{t^2}{2}\sigma^2}\nonumber \\
\frac{\dd^3}{\dd t^3}\mathcal M_{\Phi 1}(t)&=&\left(t^3\sigma^2+3t\right)\sigma^4e^{\frac{t^2}{2}\sigma^2}\,.
\end{eqnarray}
We can then calculate the characteristic function at order $k=2$:        
\begin{equation}
\mathcal M_{\Phi 2}(t)\equiv\langle e^{t\Phi}\rangle_2=\left(1+\frac{1}{3!}\xi_3t^3\right)e^{\frac{t^2}{2}\sigma^2}\,.
\end{equation}
Defining the integral over the skewness term by:
\begin{equation}
\mathcal S_t\equiv\frac{1}{3!}\xi_3t^3\,,
\end{equation}
we can write the second order moment of the density as:
\begin{equation}
\langle\rho^t\rangle_2=\overline{\rho}^te^{t\mu_s}\langle e^{t\Phi}\rangle_2=\overline{\rho}^te^{t\mu_s+\frac{t^2}{2}\sigma^2}\left(1+S_t\right)\,.
\end{equation}

The mean and variance can be obtained by evaluating the latter expression at $t=1$ and $t=2$ respectively.

\subsubsection{Mean}

\begin{equation}
\mu_s=-\frac{1}{2}\sigma^2-\ln\left(1+\mathcal S_1\right)\,.
\end{equation}

\subsubsection{Variance}

\begin{equation}
\sigma^2=\ln\left[\left(1+\mathcal S_1\right)\left(1+\mathcal S_2\right)^{-1}\left(\sigma_\delta^2+1\right)\right]\,.
\end{equation}

\subsection{Case $k=3$: lognormal with skewness and kurtosis}
\label{app:k3}

Here we include also the kurtosis terms:
\begin{eqnarray}
\lefteqn{\mathcal M_{\Phi 3}(t)\equiv\langle e^{t\Phi}\rangle_3}\\
&&\hspace{-1.cm}=\frac{1}{\sqrt{2\pi}\sigma}\int\dd \Phi\,e^{-\frac{1}{2}\frac{\Phi^2}{\sigma^2}}\left(1+\frac{1}{3!}\xi_3\tilde{h}_3+\frac{1}{4!}\xi_4\tilde{h}_4+\frac{10}{6!}\xi_3^2\tilde{h}_6\right)e^{t\Phi}\nonumber\,,
\end{eqnarray}
Since the kurtosis terms use the fourth and sixth order Hermite polynomials we need to calculate up to the sixth order derivatives of the generating function:
\begin{eqnarray}
\hspace{-0.5cm}\frac{\dd^4}{\dd t^4}\mathcal M_{\Phi 1}(t) &=&\left(t^4\sigma^4+6t^2\sigma^2+3\right)\sigma^4e^{\frac{t^2}{2}\sigma^2} \\
\hspace{-0.5cm}\frac{\dd^5}{\dd t^5}\mathcal M_{\Phi 1}(t) &=&\left(t^5\sigma^4+10t^3\sigma^2+15t\right)\sigma^6e^{\frac{t^2}{2}\sigma^2}\nonumber \\
\hspace{-0.5cm}\frac{\dd^6}{\dd t^6}\mathcal M_{\Phi 1}(t)&=&\left(t^6\sigma^6+15t^4\sigma^4+45t^2\sigma^2+15\right)\sigma^6e^{\frac{t^2}{2}\sigma^2}\nonumber\,.
\end{eqnarray}
Defining the kurtosis integral by:
\begin{equation}
\mathcal K_t\equiv\frac{1}{4!}\xi_4t^4+\frac{10}{6!}\xi_3^2t^6\,.
\end{equation}
we can write the characteristic function as:
\begin{equation}
\mathcal M_{\Phi 3}(t)\equiv\langle e^{t\Phi}\rangle_3=\left(1+\mathcal S_t+\mathcal K_t\right)e^{\frac{t^2}{2}\sigma^2}\,.
\end{equation}
The $t$-th order moment of the density yields:
\begin{equation}
\langle\rho^t\rangle_3=\overline{\rho}^te^{t\mu_s}\langle e^{t\Phi}\rangle_3=\overline{\rho}^te^{t\mu_s+\frac{t^2}{2}\sigma^2}\left(1+\mathcal S_t+\mathcal K_t\right)\,.
\end{equation}
Inserting $t=1$ and $t=2$ yields the mean and the variance respectively.

\subsubsection{Mean}

\begin{equation}
\mu_s=-\frac{1}{2}\sigma^2-\ln\left(1+\mathcal S_1+\mathcal K_1\right)\,.
\end{equation}

\subsubsection{Variance}

\begin{equation}
\sigma^2=\ln\left[\left(1+\mathcal S_1+\mathcal K_1\right)\left(1+\mathcal S_2+\mathcal K_2\right)^{-1}\left(\sigma_\delta^2+1\right)\right]\,.
\end{equation}

\section{Multivariate mean and covariance}
\label{app:multivar}

In this section we  derive the multivariate relations for the mean and variance of the expanded lognormal distribution.
Let us define the $k$-th order characteristic function for the variable $\mbi\Phi$ as a function of $\mbi t$:
\begin{eqnarray}
\lefteqn{\mathcal M_{\mbi \Phi k}(t_{1}\dots t_{n})\equiv \sum_{q_{1}\dots q_{n}=0}^\infty \langle \Phi_{i_1}^{q_{1}}\dots \Phi_{i_n}^{q_{n}} \rangle_k  \frac{t_{1}^{q_{1}}\dots t_{n}^{q_{n}}}{q_{1}!\dots q_{n}!}}\\
&&\hspace{-0cm}=\langle \exp\left(\sum_{l} t_{l}\Phi_{i_l}\right)\rangle_k= \int \dd \mbi\Phi P_k(\mbi\Phi) \exp\left(\sum_{l} t_{l}\Phi_{i_l}\right)\nonumber\,.
\end{eqnarray} 
Let us consider now the different $k$-order to calculate the corresponding mean and variance.

\subsection{Case $k=1$: lognormal}
\label{app:mk1}

We shift the Gaussian integral to obtain the solution  to the characteristic function of the lognormal distribution:
\begin{eqnarray}
\lefteqn{\mathcal M_{\mbi\Phi 1}(t_{1}\dots t_{n})\equiv\langle \exp\left(\sum_{l} t_{l}\Phi_{i_l}\right)\rangle_1}\\
&&\hspace{-.8cm}\propto\int\dd \mbi\Phi\,\exp\left({-\frac{1}{2}\sum_{i_li_m}\Phi_{i_l}S_{i_li_m}^{-1}\Phi_{i_m}+\sum_{l} t_{l}\Phi_{i_l}}\right)\nonumber\\
&&\hspace{-.8cm}\propto\exp\left({\frac{1}{2}\sum_{lm}t_{l}S_{i_li_m}t_{m}}\right)\int\dd \mbi\Phi\,\nonumber\\
&&\hspace{-.8cm}\times\exp\left(-\frac{1}{2}\sum_{i_li_m}\left(\Phi_{i_l}-\sum_{l'}S_{i_li_{l'}}t_{l'}\right)S_{i_li_m}^{-1}\left(\Phi_{i_m}-\sum_{m'}S_{i_mi_{m'}}t_{m'}\right)\right)\nonumber\,,
\end{eqnarray}
which simplifies to:
\begin{equation}
\label{eq:m1}
\mathcal M_{\mbi\Phi 1}(t_{1}\dots t_{n})\equiv\langle \exp\left(\sum_{l} t_{l}\Phi_{i_l}\right)\rangle_1=\exp\left({\frac{1}{2}\sum_{lm}t_{l}S_{i_li_m}t_{m}}\right)\nonumber\,.
\end{equation}

Taking into account that the density field $\mbi\rho$ is related to $\mbi\Phi$ by: $\rho_i=\overline{\rho}\exp\left( \Phi_i+\mu_{si}  \right)$,
we get:
\begin{equation}
{\langle\rho_{i_1}\dots\rho_{i_n}\rangle_1=}\overline{\rho}^n\exp\left(\sum_{l} t_{l}\mu_{si_l}+\frac{1}{2}\sum_{lm}t_{l}S_{i_li_m}t_{m}\right)\nonumber\,.
\end{equation}

\subsubsection{Mean}

Setting $l=1$ and  $t_1=1$ yields the mean:
\begin{equation}
\mu_{si}=-\frac{1}{2}S_{ii}\,.
\end{equation}

\subsubsection{Covariance}

Considering now $l=1,2$ and  $t_1,t_2=1$ we obtain the second moment of $\rho$:
\begin{equation}
{\langle\rho_{i}\rho_{j}\rangle_1=}
\overline{\rho}^2\exp\left(\mu_{si}+\mu_{sj}+\frac{1}{2}\left(S_{ii}+S_{jj}+S_{ij}+S_{ji}\right)\right)\,.
\end{equation}
which leads to the covariance:
\begin{equation}
S_{ij}=\ln\left(\langle\delta_{{\rm M}i}\delta_{{\rm M}j}\rangle_1+1\right)\,.
\end{equation}

\subsection{Case $k=2$: lognormal with skewness}
\label{app:mk2}

Let us recall the expression for the skewness (Eq.~\ref{eq:skew}):
\begin{equation}
\label{eq:skewap} 
\mathcal S (\mbi\nu)= \frac{1}{3!}\sum_{ijk}\kappa_{ijk}{h}_{ijk}(\mbi \nu)=\frac{1}{3!}\sum_{ijk}\xi_{ijk}\tilde{h}_{ijk}(\mbi \nu)\,.
\end{equation}
The  characteristic function including skewness yields:
\begin{eqnarray}
\lefteqn{\mathcal M_{\mbi\Phi 2}(t_{1}\dots t_{n})\equiv\langle e^{\sum_{l} t_{l}\Phi_{i_l}}\rangle_2}\\
&&\hspace{-0.5cm}\propto\int\dd \mbi\Phi\,\exp\left({-\frac{1}{2}\sum_{i_li_m}\Phi_{i_l}S_{i_li_m}^{-1}\Phi_{i_m}+\sum_{l} t_{l}\Phi_{i_l}}\right)\nonumber\\
&&\hspace{-0.5cm}\times\left(1+\frac{1}{3!} \sum_{ijk}\xi_{ijk}\tilde{h}_{ijk}(\mbi \nu) \right)\nonumber\,.
\end{eqnarray}
This integral can be solved by calculating the different moments of the lognormal generating function:
\begin{eqnarray}
\label{eq:dm1}
\hspace{-1.5cm}\lefteqn{\langle e^{\sum_{l} t_{l}\Phi_{i_l}}\Phi_{i_1}\dots\Phi_{i_n}\rangle_1=\frac{\partial^n}{\partial t_{1}\dots\partial t_{n}}\mathcal M_{\mbi\Phi 1}(t_{1}\dots t_{n})\propto}\\
&&\hspace{-.5cm}\int\dd \mbi\Phi\,\exp\left({-\frac{1}{2}\sum_{i_li_m}\Phi_{i_l}S_{i_li_m}^{-1}\Phi_{i_m}+\sum_{l} t_{l}\Phi_{i_l}}\right)\Phi_{i_1}\dots\Phi_{i_n}\nonumber\,.
\end{eqnarray}
Performing the derivatives we get:
\begin{eqnarray}
\frac{\partial}{\partial t_1}\mathcal M_{\mbi\Phi 1}(\mbi t)&=& \sum_lS_{i_1i_l}t_l\exp\left({\frac{1}{2}\sum_{l'm'}t_{l'}S_{i_{l'}i_{m'}}t_{m'}}\right) \nonumber\\
\frac{\partial^2}{\partial t_1\partial t_2}\mathcal M_{\mbi\Phi 1}(\mbi t)&=& \left(S_{i_1i_2} +\sum_lS_{i_1i_l}t_l\sum_{m}S_{i_2i_{m}}t_{m}\right) \nonumber\\
&&\times\exp\left({\frac{1}{2}\sum_{l'm'}t_{l'}S_{i_{l'}i_{m'}}t_{m'}}\right)\nonumber\\
\frac{\partial^3}{\partial t_1\partial t_2\partial t_3}\mathcal M_{\mbi\Phi 1}(\mbi t)&=& \left(S_{i_1i_2}\sum_lS_{i_3i_l}t_l +S_{i_2i_{3}}\sum_lS_{i_1i_l}t_l\right.\nonumber\\
&&\hspace{-3cm}\left.+S_{i_1i_3}\sum_{l}S_{i_2i_{l}}t_{l}+\sum_lS_{i_1i_l}t_l\sum_{m}S_{i_2i_{m}}t_{m}\sum_nS_{i_3i_n}t_n\right) \nonumber\\
&&\times\exp\left({\frac{1}{2}\sum_{l'm'}t_{l'}S_{i_{l'}i_{m'}}t_{m'}}\right)\,.
\end{eqnarray}
This leads to the $n$-order moment of the density field $\mbi\rho$:
\begin{eqnarray}
\lefteqn{\langle\rho_{i_1}\dots\rho_{i_n}\rangle_2=\overline{\rho}^n\exp\left(\sum_{l'} t_{l'}\mu_{si_{l'}}+\frac{1}{2}\sum_{l'm'}t_{l'}S_{i_{l'}i_{m'}}t_{m'}\right)}\nonumber\\
&&\hspace{-0cm}\times\left(1+ \frac{1}{3!} \sum_{i_{1}'i_{2}'i_{3}'}\xi_{i_{1}'i_{2}'i_{3}'} \right.\\
&&\hspace{-0cm}\left.\times\sum_{i_1i_2i_3}S_{i_1i_1'}^{-1}S_{i_2i_2'}^{-1}S_{i_3i_3'}^{-1} \sum_lS_{i_1i_l}t_l\sum_{m}S_{i_2i_{m}}t_{m}\sum_nS_{i_3i_n}t_n\right) \nonumber\,,
\end{eqnarray}
which can be simplified to:
\begin{eqnarray}
\label{eq:momskew}
\lefteqn{\langle\rho_{i_1}\dots\rho_{i_n}\rangle_2=\overline{\rho}^n\exp\left(\sum_{l'} t_{l'}\mu_{si_{l'}}+\frac{1}{2}\sum_{l'm'}t_{l'}S_{i_{l'}i_{m'}}t_{m'}\right)}\nonumber\\
&&\hspace{-0cm}\times\left(1+ \frac{1}{3!} \sum_{lmn}\xi_{i_{l}i_{m}i_{n}}t_lt_mt_n     \right) \,.
\end{eqnarray}

\subsubsection{Mean}

By setting  $l,m,n=1$ and $t_1=1$ we then obtain the mean:
\begin{equation}
\mu_{si}=-\frac{1}{2}S_{ii}-\ln\left(1+\frac{1}{3!}\xi_{iii}\right)\,.
\end{equation}

\subsubsection{Covariance}

Substituting  $l,m,n=1,2$ together with $t_1,t_2=1$ in Eq.~(\ref{eq:momskew})  we get the second order moment of the density:
\begin{eqnarray}
\lefteqn{\langle\rho_{i}\rho_{j}\rangle_2=
\overline{\rho}^2\exp\left(S_{ij}\right)\left(1+\frac{1}{3!}\xi_{iii}\right)^{-1}}\\
&&\hspace{-1cm}\times\left( 1+\frac{1}{3!}\left( \xi_{iii}+\xi_{jjj}+\xi_{iij}+\xi_{iji}+\xi_{jii}+\xi_{ijj}+\xi_{jij}+\xi_{jji}  \right) \right)\nonumber\,,
\end{eqnarray}
which gives us the expression for the covariance:
\begin{eqnarray}
\lefteqn{S_{ij}=\ln\left(\langle\delta_{{\rm M}i}\delta_{{\rm M}j}\rangle_2+1\right)+\ln\left(1+\frac{1}{3!}\xi_{iii}\right)}\\
&&\hspace{-1.cm}-\ln\left(1+\frac{1}{3!}\left( \xi_{iii}+\xi_{jjj}+\xi_{iij}+\xi_{iji}+\xi_{jii}+\xi_{ijj}+\xi_{jij}+\xi_{jji} \right)\right)\nonumber\,.
\end{eqnarray}
 Please note, that contracting the expressions found here to the univariate case gives the same relations as found in the previous section.

\subsection{Case $k=3$: lognormal with skewness and kurtosis}
\label{app:mk3}

The first contribution to the kurtosis term $\mathcal K$ is given by (see Eq.~\ref{eq:edge}): 
\begin{equation}
 \hspace{-0.cm}\mathcal K_{\rm A} (\mbi\nu)\equiv\frac{1}{4!}\sum_{ijkl}\kappa_{ijkl}{h}_{ijkl}(\mbi \nu)=\frac{1}{4!}\sum_{ijkl}\xi_{ijkl}\tilde{h}_{ijkl}(\mbi \nu)\,,
\end{equation}
and the second contribution is: 
\begin{eqnarray} 
\mathcal K_{\rm B} (\mbi\nu)&\equiv&\\
&&\hspace{-2.5cm} \frac{1}{6!}\sum_{i_1\dots i_6}\left[\frac{1}{3!3!2}\sum_{j_1\dots j_6\in[1,\dots,6]}\tilde{\epsilon}_{j_1\dots j_6}\xi_{i_{j_1}i_{j_2}i_{j_3}}\xi_{i_{j_4}i_{j_5}i_{j_6}}\right]_{10}\tilde{h}_{i_1\dots i_6}(\mbi \nu)\nonumber\,.
\end{eqnarray}

The characteristic function including skewness and kurtosis yields:
\begin{eqnarray}
\lefteqn{\mathcal M_{\mbi\Phi 3}(t_{1}\dots t_{n})\equiv\langle e^{\sum_{l} t_{l}\Phi_{i_l}}\rangle_3}\\
&&\hspace{-0cm}\propto\int\dd \mbi\Phi\,\exp\left({-\frac{1}{2}\sum_{i_li_m}\Phi_{i_l}S_{i_li_m}^{-1}\Phi_{i_m}+\sum_{l} t_{l}\Phi_{i_l}}\right)\nonumber\\
&&\hspace{-0cm}\times\left(1+\mathcal S(\mbi \nu)+\mathcal K(\mbi \nu) \right)\nonumber\,,
\end{eqnarray}

We can then formulate the $n$-order moment of the density field:
\begin{eqnarray}
\label{eq:nmomrho}
\lefteqn{\hspace{-0.cm}\langle\rho_{i_1}\dots\rho_{i_n}\rangle_2=\overline{\rho}^n\exp\left(\sum_{l'} t_{l'}\mu_{si_{l'}}+\frac{1}{2}\sum_{l'm'}t_{l'}S_{i_{l'}i_{m'}}t_{m'}\right)}\\
&&\hspace{-1.2cm}\times\left(1+ \frac{1}{3!} \sum_{j_{1}j_{2}j_{3}}\xi_{i_{j_1}i_{j_2}i_{j_3}}t_{j_1}t_{j_2}t_{j_3}   + \frac{1}{4!} \sum_{j_{1}\dots j_4}\xi_{i_{j_1}\dots i_{j_4}}t_{j_1}\dots t_{j_4}     \right.\nonumber\\
&&\hspace{-1.2cm}+\left. \frac{1}{6!}\sum_{j_{1}\dots j_6}\left[\frac{1}{3!3!2}\sum_{k_1\dots k_6\in[1,\dots,6]}\tilde{\epsilon}_{k_1\dots k_6}\xi_{ i_{j_{k_1}} i_{j_{k_2}} i_{j_{k_3}} } \xi_{ i_{j_{k_4}} i_{j_{k_5}} i_{j_{k_6}} }\right]_{10} \right.\nonumber\\
&&\hspace{-1.2cm}\times t_{j_1}\dots t_{j_6}  \Bigg) \nonumber\,.
\end{eqnarray}

\subsubsection{Mean}

The mean is obtained by setting $j_1,\dots,j_6=1$ and $t_1=1$:
\begin{equation}
\mu_{si}=-\frac{1}{2}S_{ii}-\ln\left(1+\frac{1}{3!}\xi_{iii}+\frac{1}{4!}\xi_{iiii}+\frac{10}{6!}\xi_{iii}^2\right)\,.
\end{equation}

\subsubsection{Covariance}
\label{sec:covkurt}

From Eq.~(\ref{eq:nmomrho}) we get the second order moment by inserting $j_1,\dots,j_6=1,2$ and $t_1,t_2=1$:
\begin{eqnarray}
\lefteqn{\langle\rho_{i}\rho_{j}\rangle_2=
\overline{\rho}^2\exp\left(S_{ij}\right)\left(1+\frac{1}{3!}\xi_{iii}+\frac{1}{4!}\xi_{iiii}+\frac{10}{6!}\xi_{iii}^2\right)^{-1}}\\
&&\times\left( 1+\frac{1}{3!}T_{\mathcal S}+\frac{1}{4!}T_{\mathcal K_{\rm A}} +\frac{1}{6!}T_{\mathcal K_{\rm B}}\right)\nonumber\,,
\end{eqnarray}
with the following definitions:
\begin{eqnarray}
T_{\mathcal S}&\equiv&\xi_{iii}+\xi_{jjj}+\xi_{iij}+\xi_{iji}+\xi_{jii}+\xi_{ijj}+\xi_{jij}+\xi_{jji}\nonumber\\
T_{\mathcal K_{\rm A}}&\equiv& \xi_{iiii}+\xi_{jjjj}+\xi_{iiij}+\xi_{iiji}+\xi_{ijii}+\xi_{jiii}+\xi_{jjji}\nonumber\\
&&  +\xi_{jjij}+\xi_{jijj}+\xi_{ijjj}+\xi_{iijj}+\xi_{ijij}\nonumber\\
&& +\xi_{ijji} +\xi_{jiij}+\xi_{jiji}+\xi_{jjii}\,.
\end{eqnarray}

\renewcommand{\arraystretch}{1.5}

\begin{table}
\begin{tabular}{c|c|c}
$(j_1|j_2|j_3|j_4|j_5|j_6)$&number of terms&configurations\\ \cline{1-3}
$(1|1|1|1|1|1)$&1=6!/6!&$\xi_{iii}\xi_{iii}$\\ \cline{1-3}
$(1|1|1|1|1|2)$&6=6!/5!&$\xi_{iii}\xi_{iij}|_{6=2\cdot3}$\\ \cline{1-3}
$(1|1|1|1|2|2)$&15=6!/(4!2)&$\xi_{iii}\xi_{ijj}|_{6=2\cdot3}$,$\xi_{iij}\xi_{iij}|_{9=3\cdot3}$\\ \cline{1-3}
$(1|1|1|2|2|2)$&20=6!/(3!3!)&$\xi_{iii}\xi_{jjj}|_{2}$,$\xi_{iij}\xi_{ijj}|_{18=2\cdot3\cdot3}$\\ \cline{1-3}
$(1|1|2|2|2|2)$&15=6!/(4!2)&$\xi_{jjj}\xi_{jii}|_{6=2\cdot3}$,$\xi_{jji}\xi_{jji}|_{9=3\cdot3}$\\ \cline{1-3}
$(1|2|2|2|2|2)$&6=6!/5!&$\xi_{jjj}\xi_{jji}|_{6=2\cdot3}$\\ \cline{1-3}
$(2|2|2|2|2|2)$&1=6!/6!&$\xi_{jjj}\xi_{jjj}$\\ \cline{1-3}
\end{tabular}
\caption{Configurations of $\xi_{i_{j_1}i_{j_2}i_{j_3}}\xi_{i_{j_4}i_{j_5}i_{j_6}}$ for $j_1,\dots,j_6=1,2$ as needed for the kurtosis contribution to the covariance under certain symmetry conditions (see section \ref{sec:covkurt}). In the first column we consider the different configurations of the indices $i\equiv i_1$ and $j\equiv i_2$ disregarding the position in the correlation functions. In the second column we calculate the number of permutations for each configuration disregarding the position of the indices. In the third column we take into account the position of the indices and identify the different classes of configurations. Note that the sum of all the terms gives 64 as we expected. } 
\label{fig:conf}
\end{table}

Please note that $T_{\mathcal K_{\rm B}}$ can be obtained in an analogous way from the last term in  Eq.~(\ref{eq:nmomrho}) and has $10\times2^6=640$ terms.
However, the number of different classes of terms can be drammatically reduced by assuming certain symmetries. In particular assuming that any permutation of the indices $\xi_{iij}$ and the permutation $i\rightarrow j$ gives identical terms as it is done in the hierarchical model, leads to only 9 classes of terms (see Tab.~\ref{fig:conf}): $\xi_{iii}^2|_{2}$, $\xi_{iii}\xi_{iij}|_{12=2\cdot2\cdot3}$, $\xi_{iii}\xi_{ijj}|_{12=2\cdot2\cdot3}$, $\xi_{iij}\xi_{iij}|_{18=2\cdot3\cdot3}$, $\xi_{iii}\xi_{jjj}|_{2}$, $\xi_{iij}\xi_{ijj}|_{18=2\cdot3\cdot3}$ with all together  64 terms which multiplied by a factor 10 give 640 terms. 
The particular expression we find has then the following form:
\begin{equation}
T_{\mathcal K_{\rm B}}\equiv20\left( 2\xi_{iii}^2+6\xi_{iii}\xi_{iij}+6\xi_{iii}\xi_{ijj}+9\xi_{iij}\xi_{iij}+9\xi_{iij}\xi_{ijj}\right)\,.
\end{equation}

The covariance is given by:
\begin{eqnarray}
\lefteqn{S_{ij}=\ln\left(\langle\delta_{{\rm M}i}\delta_{{\rm M}j}\rangle_2+1\right)+\ln\left(1+\frac{1}{3!}\xi_{iii}+\frac{1}{4!}\xi_{iiii}+\frac{10}{6!}\xi_{iii}^2\right)}\nonumber\\
&&-\ln\left(1+\frac{1}{3!}T_{\mathcal S}+\frac{1}{4!}T_{\mathcal K_{\rm A}} +\frac{1}{6!}T_{\mathcal K_{\rm B}} \right)\,.
\end{eqnarray}

\section{Multivariate skewness terms in the hierarchical model}
\label{app:skew}

Let us recall the expression for the skewness (Eq.~\ref{eq:skew}) we want to calculate here:
\begin{eqnarray}
\label{eq:skewap} 
\mathcal S (\mbi\nu)&=& \frac{1}{3!}\sum_{ijk}\kappa_{ijk}{h}_{ijk}(\mbi \nu)=\frac{1}{3!}\sum_{ijk}\xi_{ijk}\tilde{h}_{ijk}(\mbi \nu)\\
&&\hspace{-1cm}=\frac{Q_3}{3!}\sum_{ijk}\left[S_{ij}S_{ik}+S_{ij}S_{jk}+S_{ik}S_{jk}\right]\tilde{h}_{ijk}(\mbi \nu)\nonumber\,,
\end{eqnarray}
where we have inserted the three-point correlation function from the hierarchical model.

In the following subsection we will calculate the contribution of each Hermite polynomial term separately.

\subsection{First Hermite term}

Here we go through all the correlation terms applied to the first Hermite term: 
\begin{equation}
\left.
\begin{array}{l l}
\sum_{ijk}S_{ij}S_{ik}\eta_{i}\eta_{j}\eta_{k}&\hspace{-0.3cm}=\sum_{i}\eta_{i}\Phi_{i}^2\\
\sum_{ijk}S_{ij}S_{jk}\eta_{i}\eta_{j}\eta_{k}&\hspace{-0.3cm}=\sum_{j}\eta_{j}\Phi_{j}^2\\
\sum_{ijk}S_{ik}S_{jk}\eta_{i}\eta_{j}\eta_{k}&\hspace{-0.3cm}=\sum_{k}\eta_{k}\Phi_{k}^2
\end{array}\right\}3\sum_{i}\Phi_{i}^2\eta_{i} \,,
\end{equation}
with the factor 3 being due to the fact that the result is identical for the three terms of the three-point correlation function. 

\subsection{Rest of Hermite terms}

Let us look at the case in which the index of $\eta$ is the same as the one which is doubly present in the correlation term $S_{ij}S_{ik}$: 
\begin{equation}
\label{eq:three1}
3\times\sum_{ijk}S_{ij}S_{ik}\eta_{i}S^{-1}_{jk}=3\times\sum_{ii}S_{ii}\eta_{i} \,.
\end{equation}
This occurs once for each index.
Alternatively, the index of $\eta$ coincides with one of the other two indices:
\begin{equation}
\left.
\begin{array}{l l}
\label{eq:three2}
3\times\sum_{ijk}S_{ij}S_{jk}\eta_{i}S^{-1}_{jk}&\hspace{-0.3cm}=3\times\sum_{ij}S_{ij}\eta_{i}\\
3\times\sum_{ijk}S_{ik}S_{jk}\eta_{i}S^{-1}_{jk}&\hspace{-0.3cm}=3\times\sum_{ik}S_{ik}\eta_{i}
\end{array}\right\}6\sum_{i}\Phi_{i} \,.
\end{equation}
This happens twice.
The factor 3 in Eqs.~(\ref{eq:three1},\ref{eq:three2}) stands for the different leafs/labellings combined with the Hermite terms.

\subsection{Result}

Hence, the skewness term is given by:
\begin{eqnarray}
\label{eq:skewres}
\lefteqn{\mathcal S(\mat S^{-1/2}\mbi \Phi)}\\
&&\hspace{-.5cm}=Q_3\left[\frac{1}{2}\sum_{i}\Phi_{i}^2\eta_{i}-\frac{1}{2}\sum_{i}S_{ii}\eta_{i}- \sum_{i}\Phi_{i}\right]\nonumber\\
&&\hspace{-.5cm}=Q_3\left[\frac{1}{2}\sum_{i}\Phi_{i}^2\sum_{j}S_{ij}^{-1}\Phi_{j}-\frac{1}{2}\sum_{i}S_{ii}\sum_{j}S_{ij}^{-1}\Phi_{j}-\sum_{i}\Phi_{i} \right]\nonumber\,.
\end{eqnarray}

\section{Multivariate kurtosis terms in the hierarchical model}
\label{app:kurt}

The Edgeworth expansion shows that there are two groups of terms contributing to the third order perturbation, which we call the  kurtosis:
$\mathcal K\equiv\mathcal K_{\rm A}+\mathcal K_{\rm B} $.
Let us look at each group separately.

\subsection{First group of kurtosis terms}
\label{app:kurt1}

The first contribution $\mathcal K_{\rm A}$  to the kurtosis term $\mathcal K$ is given by (see Eq.~\ref{eq:edge}): 
\begin{eqnarray}
 \hspace{-.0cm}\mathcal K_{\rm A} (\mbi\nu)&\equiv& \frac{1}{4!}\sum_{ijkl}\kappa_{ijkl}{h}_{ijkl}(\mbi \nu)=\frac{1}{4!}\sum_{ijkl}\xi_{ijkl}\tilde{h}_{ijkl}(\mbi \nu)\nonumber\,
\end{eqnarray}
or more specifically:
\begin{eqnarray} 
\mathcal K_{\rm A} (\mbi\nu)&\equiv&\\
&&\hspace{-.5cm}Q_4^{\rm a}\left[ S_{ij}S_{ik}S_{il}+S_{ij}S_{jk}S_{jl}+S_{ik}S_{jk}S_{kl}+S_{il}S_{jl}S_{kl} \right]\nonumber\\
&&\hspace{-.5cm}+Q_4^{\rm b}\left[S_{ij}S_{ik}S_{jl}+S_{ij}S_{il}S_{jk}+S_{ij}S_{ik}S_{kl}+S_{il}S_{ik}S_{kj}  \right.\nonumber\\
&&\hspace{-.5cm}\left.+ S_{ij}S_{il}S_{kl}+S_{ik}S_{il}S_{jl}+S_{ij}S_{jk}S_{kl}+S_{ik}S_{jk}S_{jl}\right.\nonumber\\
&&\hspace{-.5cm}\left.+S_{ij}S_{jl}S_{kl}+S_{il}S_{jk}S_{jl}+S_{ik}S_{jl}S_{kl}+S_{il}S_{jk}S_{kl} \right]  \nonumber\\
&&\hspace{-.5cm}
\times\tilde{h}_{ijkl}(\mbi \nu)\nonumber\,,
\end{eqnarray}
where we have inserted the four-point correlation function from the hierarchical model.

Since the four-point correlation function has two trees (see section \ref{sec:HM}) we will calculate the terms for each tree separately.

\subsubsection{First tree}

The terms corresponding to the first tree are:
\begin{enumerate}
\item {\it  First Hermite term}: there are 3 indices for the 3 $\eta$'s of the Hermite polynomial singly coupled to the corresponding index in the $S$ functions (in the case chosen here: $j,k,l$). One index remains for the final contraction of the term (here: $i$):
\begin{equation}
\hspace{0cm}4\times\sum_{ijkl}S_{ij}S_{ik}S_{il}\eta_{i}\eta_{j}\eta_{k}\eta_{l}=4\times\sum_{i}\eta_{i}\Phi_{i}^3 \,.
\end{equation}
The factor 4 comes from the fact that this will occur for the 4 indices which are run by the different leafs of this tree of the four-point correlation function. 
\item {\it  Second Hermite term}: 2 indices are assigned to 2 $\eta$'s singly coupled to the corresponding index in the $S$ functions  (for example: $j,k$). 2 remaining indices are assigned to $S^{-1}$ with 1 index singly coupled to the corresponding index in $S$ (for example: $l$) and 1 index triply coupled (for example: $i$). This can only happen thrice for each term of the four-point correlation function and 4 times for each of the indices ($4\times3$):
\begin{equation}
\hspace{0cm}4\times3\times\sum_{ijkl}S_{ij}S_{ik}S_{il}\eta_{j}\eta_{k}S^{-1}_{il}=4\times3\times\sum_{i}\Phi^2_{i}\,.
\end{equation}
The other possibility for this Hermite term is that 1 index of 2 for the 2 $\eta$'s is singly coupled to the corresponding index in $S$  and the other one is triply coupled. The other 2 indices are assigned to $S^{-1}$ (with occurrence $4\times3$):
\begin{equation}
\hspace{0cm}4\times3\times\sum_{ijkl}S_{ij}S_{ik}S_{il}\eta_{i}\eta_{j}S^{-1}_{kl}=4\times3\times\sum_{i}S_{ii}\eta_{i}\Phi_{i} 
 \,.
\end{equation}

\item {\it  Third Hermite term}:
The 2 indices of one of $S^{-1}$ (for example: $k,l$) are singly coupled to the corresponding indices of the $S$ functions. The indices of the remaining $S^{-1}$ will be one singly coupled (for example: $j$) and 1 triply coupled (for example: $i$). This happens 3 times for each four-point correlation term:
\begin{equation}
\hspace{0cm}4\times3\times\sum_{ijkl}S_{ij}S_{ik}S_{il}S^{-1}_{ij}S^{-1}_{kl}=4\times3\times\sum_{i}S_{ii} \,.
\end{equation}

\end{enumerate}

\subsubsection{Second tree}

The terms corresponding to the second tree are:

\begin{enumerate}
\item {\it First Hermite term}:
the only possible configuration for each four-point correlation term is that 2 $\eta$'s are singly coupled (for example: $k,l$) and 2 $\eta$'s are doubly coupled to the $S$ functions (for example: $i,j$):
\begin{equation}
\hspace{0cm}12\times\sum_{ijkl}S_{ij}S_{ik}S_{jl}\eta_{i}\eta_{j}\eta_{k}\eta_{l}=12\times\sum_{ij}\eta_{i}\Phi_{i}S_{ij}\Phi_{j}\eta_{j} \,,
\end{equation}
thus having a factor 12 for all leafs.

\item {\it  Second Hermite term}:
2 $\eta$'s are singly coupled to 2 $S$ functions (for example with indices: $k,l$). The remaining indices for $S^{-1}$ are doubly coupled (for example: $i,j$). This happens only once for each four-point correlation term:
\begin{equation}
 \hspace{0cm}12\times\sum_{ijkl}S_{ij}S_{ik}S_{jl}\eta_{k}\eta_{l}S^{-1}_{ij}=12\times\left(\sum_{i}\Phi_{i}\right)^2 
\,.
\end{equation}
Let us consider now 1 $\eta$ is singly coupled (for example: $k$ or $l$) and the other $\eta$ is doubly coupled (for example: $i$ or $j$) to the $S$ functions. There are two possibilities: the singly coupled index appears in the same $S$ function with the doubly coupled index:  
\begin{equation}
\hspace{0cm}12\times2\times\sum_{ijkl}S_{ij}S_{ik}S_{jl}\eta_{i}\eta_{k}S^{-1}_{jl}=12\times2\times\sum_{ij}S_{ij}\eta_{i}\Phi_{i} \,,
\end{equation}
or the singly coupled index appears in the remaining $S$ function:  
\begin{equation}
\hspace{0cm}12\times2\times\sum_{ijkl}S_{ij}S_{ik}S_{jl}\eta_{i}\eta_{l}S^{-1}_{jk}=12\times2\times\sum_{ij}S_{ii}\eta_{i}\Phi_{i} \,.
\end{equation}
There are 2 configurations for both cases for each four-point correlation term, hence $12\times2$.

The last configuration for the second Hermite term is based on 2 indices of $\eta$ doubly coupled to the $S$ functions (this happens only once for each four-point correlation term: 12 times in total):
\begin{equation}
\hspace{0cm}12\times\sum_{ijkl}S_{ij}S_{ik}S_{jl}\eta_{i}\eta_{j}S^{-1}_{kl}=12\times\sum_{ij}\eta_{i}S^2_{ij}\eta_{j}\,,
\end{equation}

\item {\it  Third Hermite term}:
the last Hermite term has two possibilities: 
either only one of the indices of the $S^{-1}$ functions coincides respectively with one of the indices of the $S$ functions (this can only happen once for each four-point correlation term):
\begin{equation}
\hspace{0cm}12\times\sum_{ijkl}S_{ij}S_{ik}S_{jl}S^{-1}_{il}S^{-1}_{jk}=12\times\sum_{i}S_{ii} \,,
\end{equation}
or two indices of one of the $S^{-1}$ functions coincide with the corresponding indices of one of the $S$ functions (this can occur twice for each four-point correlation term: $12\times2$):  
\begin{equation}
\hspace{0cm}12\times2\times\sum_{ijkl}S_{ij}S_{ik}S_{jl}S^{-1}_{ij}S^{-1}_{kl}=12\times2\times\sum_{ij}S_{ij} \,,
\end{equation}

\end{enumerate}

\subsubsection{Result}

Putting the contribution of both trees together we get the first kurtosis term:
\begin{eqnarray}
\lefteqn{\mathcal K_{\rm A}(\mat S^{-1/2}\mbi \Phi)=}\\
&&\hspace{-0.5cm}\frac{Q_4^{\rm a}}{2}\left[ \frac{1}{3}\sum_{i}\Phi_{i}^3\eta_i-\sum_{i}\Phi^2_{i}\right]\nonumber\\
&&\hspace{-0.5cm}-\left(\frac{Q_4^{\rm a}}{2}+Q_4^{\rm b}\right) \sum_{i}S_{ii}\Phi_{i}\eta_{i}  +\frac{1}{2}\left(Q_4^{\rm a}+Q_4^{\rm b}\right)\sum_{i}S_{ii}\nonumber\\
&&\hspace{-0.5cm} +\frac{Q_4^{\rm b}}{2}\left[\sum_{ij}\eta_{i}\Phi_{i}S_{ij}\Phi_{j}\eta_{j}-\sum_{ij}\eta_{i}S^2_{ij}\eta_{j}\right.\nonumber\\
&&\hspace{-0.5cm} \left.-\left(\sum_{i}\Phi_{i}\right)^2-{2}\sum_{ij}\Phi_{i}S_{ij}\eta_{j}+2\sum_{ij}S_{ij}\right]  \nonumber\,,
\end{eqnarray}
or equivalently:
\begin{eqnarray}
\label{eq:kurt1res}
\lefteqn{\mathcal K_{\rm A}(\mat S^{-1/2}\mbi \Phi)=}\\
&&\hspace{-0.5cm}\frac{Q_4^{\rm a}}{2}\left[ \frac{1}{3}\sum_{i}\Phi_{i}^3\sum_j S_{ij}^{-1} \Phi_j-\sum_{i}\Phi^2_{i}\right]\nonumber\\
&&\hspace{-0.5cm} -\left(\frac{Q_4^{\rm a}}{2}+Q_4^{\rm b}\right) \sum_{i}S_{ii}\Phi_{i}\sum_j S_{ij}^{-1} \Phi_j  +\frac{1}{2}\left(Q_4^{\rm a}+Q_4^{\rm b}\right)\sum_{i}S_{ii}\nonumber\\
&&\hspace{-0.5cm} +\frac{Q_4^{\rm b}}{2}\left[\sum_{ij}\sum_{i} S_{ii}^{-1} \Phi_{i}\Phi_{i}S_{ij}\Phi_{j}\sum_{j} S_{jj}^{-1} \Phi_{j}\right.\nonumber\\
&&\hspace{-0.5cm}\left.-\sum_{ij}\sum_{i}S^{-1}_{ii}\Phi_{i}S^2_{ij}\sum_{j}S^{-1}_{jj}\Phi_{j}\right.\nonumber\\
&&\hspace{-0.5cm} \left.-\left(\sum_{i}\Phi_{i}\right)^2-{2}\sum_{ij}\Phi_{i}S_{ij}\sum_{j} S_{jj}^{-1} \Phi_{j}+2\sum_{ij}S_{ij}\right] \nonumber\,.
\end{eqnarray}

\subsection{Second group of kurtosis terms}

\label{app:kurt2}

The second contribution $\mathcal K_{\rm B}$  to the kurtosis term $\mathcal K$ is given by (see Eq.~\ref{eq:edge}): 
\begin{eqnarray} 
\mathcal K_{\rm B} (\mbi\nu)&\equiv& \frac{1}{6!}\sum_{i_1\dots i_6}\\
&&\hspace{-2.cm}\times\left[\frac{1}{3!3!2}\sum_{j_1\dots j_6\in[1,\dots,6]}\tilde{\epsilon}_{j_1\dots j_6}\langle\Phi_{i_{j_1}}\Phi_{i_{j_2}}\Phi_{i_{j_3}}\rangle_{\rm c}\langle\Phi_{i_{j_4}}\Phi_{i_{j_5}}\Phi_{i_{j_6}}\rangle_{\rm c}\right]_{10}\nonumber\\
&&\hspace{-2.cm}\times\tilde{h}_{i_1\dots i_6}(\mbi \nu)\nonumber\\
&&\hspace{-2.cm}=\frac{10}{6!}Q^2_3\sum_{ijklmn}\left[ S_{ij} S_{ik}+ S_{ij} S_{jk}+ S_{ik} S_{jk}\right]\nonumber\\
&&\hspace{-2.cm}\times\left[ S_{lm} S_{ln}+ S_{lm} S_{mn}+ S_{ln} S_{mn}\right]\nonumber\\
&&\hspace{-2.cm}\times\tilde{h}_{ijklmn}(\mbi \nu)\nonumber\,,
\end{eqnarray}
where we have inserted the three-point correlation function from the hierarchical model.

The three-point correlation function which has only one tree in the hierarchical model and the sixth order Hermite polynomial which has four terms. Let us partition the problem into these four Hermite terms taking into account the symmetries intrinsic to the hierarchical three-point correlation function.

\subsubsection{First Hermite term}

Here we always have 4 $\eta$'s singly coupled and 2 $\eta$'s doubly coupled to $S$ functions. Since we have the three-point correlation function squared  we will have $3\times3$ terms:
\begin{eqnarray}
\hspace{-.5cm}&&9\times\sum_{ijklmn}S_{ij}S_{ik}S_{lm}S_{ln}\eta_{i}\eta_{j}\eta_{k}\eta_{l}\eta_{m}\eta_{n}\nonumber\\
\hspace{-.5cm}&&=9\times\left(\sum_{i}\eta_{i}\Phi^2_{i}\right)^2 \,.
\end{eqnarray}

\subsubsection{Second Hermite term}

\begin{enumerate}
\item 4 $\eta$'s are singly coupled to the $S$ functions leave the 2 remaining indices for the $S^{-1}$ function doubly coupled to the $S$ functions   (this can happen only once for each of the 9 hierarchical correlation terms):
\begin{eqnarray}
\hspace{-.5cm}&&9\times\sum_{ijklmn}S_{ij}S_{ik}S_{lm}S_{ln}\eta_{j}\eta_{k}\eta_{m}\eta_{n}S^{-1}_{il}\nonumber\\
\hspace{-.5cm}&&=9\times\sum_{ij}\Phi^2_{i}S^{-1}_{ij}\Phi^2_{j}\,.
\end{eqnarray}

\item 3 $\eta$'s are singly coupled to the $S$ functions.
The remaining $\eta$ is doubly coupled to the $S$ functions. It can happen that the remaining $\eta$ is coupled to the $S$ functions which are also coupled to 2 $\eta$'s:  
\begin{eqnarray}
\hspace{-.5cm}&&9\times4\times\sum_{ijklmn}S_{ij}S_{ik}S_{lm}S_{ln}\eta_{j}\eta_{k}\eta_{m}\eta_{i}S^{-1}_{ln}\nonumber\\
\hspace{-.5cm}&&=9\times4\times\sum_{i}\Phi_{i}\sum_j\Phi^2_{j}\eta_{j}\,.
\end{eqnarray}
or only to 1:
\begin{eqnarray}
\hspace{-.5cm}&&9\times4\times\sum_{ijklmn}S_{ij}S_{ik}S_{lm}S_{ln}\eta_{j}\eta_{k}\eta_{m}\eta_{l}S^{-1}_{in}\nonumber\\
\hspace{-.5cm}&&=9\times4\times\sum_{i}\Phi^3_{i}\eta_{i}\,.
\end{eqnarray}

\item 2 $\eta$'s are singly coupled to the $S$ functions.
Both  singly coupled $\eta$ indices are coupled to $S$ functions which have a common index (can happen twice):  
\begin{eqnarray}
\hspace{-.5cm}&&9\times2\times\sum_{ijklmn}S_{ij}S_{ik}S_{lm}S_{ln}\eta_{i}\eta_{j}\eta_{k}\eta_{l}S^{-1}_{mn}\nonumber\\
\hspace{-.5cm}&&=9\times2\times\sum_{i}\Phi^2_{i}\eta_{i}\sum_j\eta_{j}S_{jj}\nonumber\,.
\end{eqnarray}
Both singly coupled $\eta$ indices are coupled to $S$ functions with different indices (has 4 combinations): 
\begin{eqnarray}
\hspace{-.5cm}&&9\times4\times\sum_{ijklmn}S_{ij}S_{ik}S_{lm}S_{ln}\eta_{i}\eta_{j}\eta_{m}\eta_{l}S^{-1}_{kn}\nonumber\\
\hspace{-.5cm}&&=9\times4\times\sum_{ij}\eta_{i}\Phi_{i}S_{ij}\eta_{j}\Phi_{j}\,.
\end{eqnarray}

\end{enumerate}

\subsubsection{Third Hermite term}

\begin{enumerate}
\item 2  $\eta$'s are singly coupled to the $S$ functions.  
Both  singly coupled $\eta$ indices are coupled to $S$ functions which have a common index:
\begin{eqnarray}
\hspace{-.5cm}&&9\times2\times\sum_{ijklmn}S_{ij}S_{ik}S_{lm}S_{ln}\eta_{j}\eta_{k}S^{-1}_{il}S^{-1}_{mn}\nonumber\\
 &&=9\times2\times\sum_i S_{ii}\sum_j S^{-1}_{ij}\Phi_j^2\,,
\end{eqnarray}
or by a permutation of the indices of the $S^{-1}$-functions:
\begin{eqnarray}
 &&9\times4\times\sum_{ijklmn}S_{ij}S_{ik}S_{lm}S_{ln}\eta_{j}\eta_{k}S^{-1}_{im}S^{-1}_{ln}\nonumber\\
 &&=9\times4\times\sum_i \Phi_i^2 \,.
\end{eqnarray}

\item 2  $\eta$'s are singly coupled to the $S$ functions.  
Both  singly coupled $\eta$ indices are coupled to $S$ functions which have no common index:
\begin{eqnarray}
\hspace{-.5cm}&&9\times4\times\sum_{ijklmn}S_{ij}S_{ik}S_{lm}S_{ln}\eta_{j}\eta_{m}S^{-1}_{ik}S^{-1}_{ln}\nonumber\\
 &&=9\times4\times\left(\sum_i \Phi_i\right)^2 \,,
\end{eqnarray}
or by a permutation of the indices of the $S^{-1}$-functions:
\begin{eqnarray}
&&\hspace{-.5cm}9\times8\times\sum_{ijklmn}S_{ij}S_{ik}S_{lm}S_{ln}\eta_{j}\eta_{m}S^{-1}_{il}S^{-1}_{kn}\nonumber\\
 &&=9\times8\times\left(\sum_i \Phi_i\right)^2 \,.
\end{eqnarray}

\item 1  $\eta$ is singly coupled to 1 $S$ function. 1  $\eta$ is doubly coupled to 2 $S$ functions with one of them sharing the same index as the singly coupled one. The rest of the indices which appear only once  and do not share the same doubly present index in the $S$ functions  do not mix in the $S^{-1}$ functions (4 combinations in the $\eta$ indices):
\begin{eqnarray} 
\hspace{-.5cm}&&9\times4\times\sum_{ijklmn}S_{ij}S_{ik}S_{lm}S_{ln}\eta_{i}\eta_{j}S^{-1}_{kl}S^{-1}_{mn}\nonumber\\
&&=9\times4\times\sum_i S_{ii}\eta_i\Phi_i \,.
\end{eqnarray}
The rest of the indices which appear only once  and do not share the same doubly present index in the $S$ functions are mixed in the $S^{-1}$ functions  (4 combinations in the $\eta$ indices and 2 combinations in the $S^{-1}$ indices):
\begin{eqnarray}
&&\hspace{-.5cm}9\times8\times\sum_{ijklmn}S_{ij}S_{ik}S_{lm}S_{ln}\eta_{i}\eta_{j}S^{-1}_{km}S^{-1}_{ln}\nonumber\\
&&=9\times8\times\sum_{ij} S_{ij}\eta_j\Phi_j  \,.
\end{eqnarray}

\item 1  $\eta$ is singly coupled to 1 $S$ function. 1  $\eta$ is doubly coupled to 2 $S$ functions with non of them sharing the same index as the singly coupled one. The rest of the indices which appear only once  and do not share the same doubly present index in the $S$ functions  do not mix in the $S^{-1}$ functions (4 combinations in the $\eta$ indices):
\begin{eqnarray} 
\hspace{-.5cm}&&9\times4\times\sum_{ijklmn}S_{ij}S_{ik}S_{lm}S_{ln}\eta_{j}\eta_{l}S^{-1}_{ik}S^{-1}_{mn}\nonumber\\
&&=9\times4\times\sum_i S_{ii}\eta_i\Phi_i \,.
\end{eqnarray}

The rest of the indices which appear only once  and do not share the same doubly present index in the $S$ functions  are mixed in the $S^{-1}$ functions (4 combinations in the $\eta$ indices and 2 combinations in the $S^{-1}$ indices):
\begin{eqnarray}
&&\hspace{-.5cm}9\times8\times\sum_{ijklmn}S_{ij}S_{ik}S_{lm}S_{ln}\eta_{j}\eta_{l}S^{-1}_{im}S^{-1}_{kn}\nonumber\\
&&=9\times8\times\sum_{i} S_{ii}\eta_i\Phi_i  \,.
\end{eqnarray}

\item Both  $\eta$'s are doubly coupled to the $S$ functions.   The rest of the indices which appear only once  and do not share the same doubly present index in the $S$ functions  do not mix in the $S^{-1}$ functions (only 1 combination):
\begin{eqnarray} 
\hspace{-.5cm}&&9\times\sum_{ijklmn}S_{ij}S_{ik}S_{lm}S_{ln}\eta_{i}\eta_{l}S^{-1}_{jk}S^{-1}_{mn}\nonumber\\
 &&=9\times\left(\sum_i S_{ii}\eta_i\right)^2 \,.
\end{eqnarray}
The rest of the indices which appear only once  and do not share the same doubly present index in the $S$ functions  are mixed in the $S^{-1}$ functions (2 combinations in the $S^{-1}$ indices):
\begin{eqnarray} 
&&\hspace{-.5cm}9\times2\times\sum_{ijklmn}S_{ij}S_{ik}S_{lm}S_{ln}\eta_{i}\eta_{l}S^{-1}_{jm}S^{-1}_{kn}\nonumber\\
 &&=9\times2\times\sum_{ij} \eta_iS_{ij}^2\eta_j   \,.
 \end{eqnarray}

\end{enumerate}

\subsubsection{Fourth Hermite term}

\begin{enumerate}
\item Both indices of 2 $S^{-1}$ functions are pairwise the same as the indices of 2 $S$ functions (4 combinations):  
\begin{eqnarray} \hspace{-.5cm}&&9\times4\times\sum_{ijklmn}S_{ij}S_{ik}S_{lm}S_{ln}S_{ij}^{-1}S^{-1}_{lm}S^{-1}_{kn}\nonumber\\
 &&=9\times4\times\sum_{ij} S_{ij} \,.
 \end{eqnarray}

\item Both indices of 1 $S^{-1}$ function are the same as the indices of 1 $S$ function (4 combinations):
\begin{eqnarray} \hspace{-.5cm}&&9\times4\times\sum_{ijklmn}S_{ij}S_{ik}S_{lm}S_{ln}S_{ij}^{-1}S^{-1}_{kl}S^{-1}_{mn}\nonumber\\
 &&=9\times4\times\sum_{i} S_{ii} \,.
 \end{eqnarray}

\item Non of the indices of the $S^{-1}$ functions coincides pairwise with the  indices of the $S$ functions.
The indices which are doubly present in the $S$ functions do not coincide with the indices of a single $S^{-1}$ function:  
\begin{eqnarray} \hspace{-.5cm}&&9\times4\sum_{ijklmn}S_{ij}S_{ik}S_{lm}S_{ln}S_{im}^{-1}S^{-1}_{jn}S^{-1}_{kl}\nonumber\\
 &&=9\times4\sum_{i} S_{ii} \,.
 \end{eqnarray}

\item Non of the indices of the $S^{-1}$ functions coincides pairwise with the  indices of the $S$ functions.
Both indices which are doubly present in the $S$ functions coincide with the indices of a single $S^{-1}$ function.  
The rest of the indices which appear only once  and do not share the same doubly present index in the $S$ functions  do not mix in the $S^{-1}$ functions (only 1 possibility):
\begin{eqnarray} \hspace{-.5cm}&&9\times\sum_{ijklmn}S_{ij}S_{ik}S_{lm}S_{ln}S_{il}^{-1}S^{-1}_{jk}S^{-1}_{mn}\nonumber\\
 &&=9\times\sum_{ij} S_{ii}S_{ij}^{-1}S_{jj} \,.
 \end{eqnarray}

The rest of the indices which appear only once  and do not share the same doubly present index in the $S$ functions  are mixed in the $S^{-1}$ functions (2 combinations):
\begin{eqnarray} \hspace{-.5cm}&&9\times2\times\sum_{ijklmn}S_{ij}S_{ik}S_{lm}S_{ln}S_{il}^{-1}S^{-1}_{jm}S^{-1}_{kn}\nonumber\\
 &&=9\times2\times\sum_{ij} S_{ij} \,.
 \end{eqnarray}

\end{enumerate}

\subsubsection{Result}

Summing up the terms we get:
\begin{eqnarray}
\lefteqn{\mathcal K_{\rm B}(\mat S^{-1/2}\mbi \Phi)=}\\
&&\hspace{-0.5cm} Q_3^2\left[ \frac{1}{8}\left(\sum_{i}\Phi_{i}^2\eta_i\right)^2-\frac{1}{8}\sum_{ij}\Phi^2_{i}S_{ij}^{-1}\Phi^2_{j}-\frac{1}{2}\sum_{ij}\Phi_{i}\Phi_j^2\eta_{j} \right.\nonumber\\
&&\hspace{-0.5cm}\left. - \frac{1}{2} \sum_{i}\Phi^3_{i}\eta_i  - \frac{1}{4} \sum_{ij}\Phi_i^2\eta_iS_{jj}\eta_j - \frac{1}{2}\sum_{ij}\eta_i\Phi_iS_{ij}\eta_j\Phi_j  \right.\nonumber\\
&&\hspace{-0.5cm}\left.+ \frac{1}{4} \sum_{ij}S_{ii}S_{ij}^{-1}\Phi_j^2 +\frac{1}{2}\sum_i\Phi_i^2+\frac{3}{2}\left(\sum_i\Phi_i \right)^2 \right.\nonumber\\
&&\hspace{-0.5cm} \left.+2\sum_iS_{ii}\eta_i\Phi_i+\sum_{ij}S_{ij}\eta_j\Phi_j+\frac{1}{8} \left( \sum_iS_{ii}\eta_i\right)^2 \right.  \nonumber\\
&&\hspace{-0.5cm} \left. +\frac{1}{4} \sum_{ij}\eta_iS_{ij}^2\eta_j -\frac{3}{4} \sum_{ij}S_{ij}-\sum_{i}S_{ii} -\frac{1}{8}\sum_{ij}S_{ii}S_{ij}^{-1}S_{jj}  \right]  \nonumber\,,
\end{eqnarray}
which can also be written as:
\begin{eqnarray}
\label{eq:kurt2res}
\lefteqn{\mathcal K_{\rm B}(\mat S^{-1/2}\mbi \Phi)=}\\
&&\hspace{-0.5cm} Q_3^2\left[ \frac{1}{8}\left(\sum_{i}\Phi_{i}^2\sum_{j}S_{ij}^{-1}\Phi_{j}\right)^2-\frac{1}{8}\sum_{ij}\Phi^2_{i}S_{ij}^{-1}\Phi^2_{j} \right.\nonumber\\
&&\hspace{-0.5cm}\left. -\frac{1}{2}\sum_{ij}\Phi_{i}\Phi_j^2\sum_{j}S_{jj}^{-1}\Phi_{j} - \frac{1}{2} \sum_{i}\Phi^3_{i}\sum_{j}S_{ij}^{-1}\Phi_{j}  \right.\nonumber\\
&&\hspace{-0.5cm}\left.  - \frac{1}{4} \sum_{ij}\Phi_i^2\sum_{i}S_{ii}^{-1}\Phi_{i}S_{jj}\sum_{j}S_{jj}^{-1}\Phi_{j} \right.\nonumber\\
&&\hspace{-0.5cm} \left.- \frac{1}{2}\sum_{ij}\sum_{i}S_{ii}^{-1}\Phi_{i}\Phi_iS_{ij}\sum_{j}S_{jj}^{-1}\Phi_{j}\Phi_j\right.\nonumber\\
&&\hspace{-0.5cm} \left.+ \frac{1}{4} \sum_{ij}S_{ii}S_{ij}^{-1}\Phi_j^2 +\frac{1}{2}\sum_i\Phi_i^2+\frac{3}{2}\left(\sum_i\Phi_i \right)^2\right.\nonumber\\
&&\hspace{-0.5cm} \left.+2\sum_iS_{ii}\sum_{j}S_{ij}^{-1}\Phi_{j}\Phi_i+\sum_{ij}S_{ij}\sum_{j}S_{jj}^{-1}\Phi_{j}\Phi_j \right.\nonumber\\
&&\hspace{-0.5cm} \left.+\frac{1}{8} \left( \sum_iS_{ii}\sum_{j}S_{ij}^{-1}\Phi_{j}\right)^2+\frac{1}{4} \sum_{ij}\sum_{i}S_{ii}^{-1}\Phi_{i}S_{ij}^2\sum_{j}S_{jj}^{-1}\Phi_{j} \right.  \nonumber\\
&&\hspace{-0.5cm} \left.-\frac{3}{4} \sum_{ij}S_{ij}  -\sum_{i}S_{ii} -\frac{1}{8}\sum_{ij}S_{ii}S_{ij}^{-1}S_{jj}  \right]  \nonumber\,.
\end{eqnarray}

Please note that one gets the same result as in Eqs.~(\ref{eq:skewUNI}, \ref{eq:kurt1UNI}, and \ref{eq:kurt2UNI})  by simplifying the corresponding Eqs.~(\ref{eq:skewres}, \ref{eq:kurt1res}, and \ref{eq:kurt2res}) to a single index. 
This actually demonstrates that we have gone through all the possible configurations of indices as the number of equivalent Hermite terms is recovered.

\end{document}